\documentclass{elsarticle}
\usepackage{hyperref}
\usepackage{pb-diagram}
\usepackage{amsmath, amssymb, amsthm}
\usepackage{leftidx}
\usepackage{tikz}
\usetikzlibrary{arrows, automata, shapes, calc, fadings}
\tikzset{->, auto, >=stealth'}
\tikzset{state/.style={shape=circle, draw, fill=white, initial text=,
    inner sep=.5mm, minimum size=1.5mm}}
\tikzset{state with output/.style={shape=rectangle split, rectangle
    split parts=2, draw, fill=white,
    initial text=, inner sep=1mm}}
\usetikzlibrary{matrix,arrows,decorations.pathmorphing}
\usepackage{tikz-cd}
\usepackage{mathrsfs}
\usepackage{paralist}
\usepackage{color}
\usepackage{stmaryrd}

\usepackage{changes}
\definechangesauthor[color=green!50!black]{k}

\newtheorem{thm}{Theorem}
\newtheorem{lem}[thm]{Lemma}
\newtheorem{prp}[thm]{Proposition}

\theoremstyle{definition}
\newtheorem{df}[thm]{Definition}
\newtheorem{exa}[thm]{Example}
\newtheorem{cor}[thm]{Corollary}

\theoremstyle{remark}
\newtheorem*{rem}{Remark}

\numberwithin{equation}{section}
\numberwithin{thm}{section}

\DeclareMathOperator{\id}{id}
\DeclareMathOperator{\Id}{Id}
\DeclareMathOperator*{\colim}{colim}

\def\Set{\mathbf{Set}}

\def\Cat{\mathbf{Cat}}

\def\dCat{\mathbf{dCat}}

\def\Lin{\mathbf{Lin}}

\def\Fib{\mathbf{Fib}}
\def\dFib{\mathbf{dFib}}
\def\Langs{\mathscr{L}}

\def\Z{\mathbb{Z}}
\def\N{\mathbb{N}}

\def\cC{\mathcal{C}}
\def\cD{\mathcal{D}}
\def\cE{\mathcal{E}}
\def\cF{\mathcal{F}}
\def\cG{\mathcal{G}}
\def\cI{\mathcal{I}}
\def\cJ{\mathcal{J}}

\def\cP{\mathcal{P}}
\def\cV{\mathcal{V}}

\def\sfa{\mathsf{a}}

\def\sfu{\mathsf{u}}
\def\sfv{\mathsf{v}}
\def\sfw{\mathsf{w}}

\def\sfO{\mathsf{0}}

\def\bI{\mathbf{1}}
\def\bt{\mathbf{t}}

\let\phi\varphi

\mathchardef\mhyphen="2D 

\newcommand{\Ob}{\mathsf{Ob}}
\newcommand{\Cell}{\mathsf{E}}
\newcommand{\Path}{\mathsf{Path}}

\newcommand{\TrO}{\mathsf{TrO}}

\newcommand{\Lang}{\mathsf{Lang}}

\newcommand{\Reg}{\mathsf{Reg}}
\newcommand{\Rat}{\mathsf{Rat}}

\newcommand*\intord{\dashrightarrow}
\newcommand{\sq}{\square}

\newcommand*\arrO[1]
{\xrightarrow[{\raisebox{2pt}[1pt][0pt]{\scriptsize$+$}}]{#1}}
\newcommand*\arrI[1]
{\xleftarrow[{\raisebox{2pt}[1pt][0pt]{\scriptsize$-$}}]{#1}}

\newcommand{\dcl}{{\downarrow}}

\newcommand*\idtrack[1]{\mathsf{Id}_{#1}}

\newcommand*\yoneda{\mathsf{Y}}

\newcommand*\src{\mathsf{src}}
\newcommand*\tgt{\mathsf{tgt}}
\newcommand*\subsu{\sqsubseteq}

\newcommand*\psh[1]{\mathbf{PSh}({#1})}
\newcommand*\aut[1]{\mathbf{PAut}({#1})}
\newcommand*\saut[1]{\mathbf{SAut}({#1})}
\newcommand*\tro[1]{\mathbf{TrO}({#1})}

\renewcommand*\path[1]{\mathbf{Path}({#1})}

\newcommand*\slice{{\downarrow}}

\newcommand*\ilo[3]{\vphantom{#2}_{#1}#2_{#3}}

\def\upstep{\mathbf{S}}
\def\downstep{\mathbf{T}}
\def\isostep{\mathbf{I}}
\def\vo{*}
\def\oldsq{\square}

\newcommand*\denser{\preceq}

\newcommand*\forM[1]{{#1}^+\!}
\newcommand*\bckM[1]{{#1}^-\!}
\newcommand*\fbM[1]{{#1}^\pm\!}
\newcommand*\isoM[1]{{#1}^\approx\!}

\newcommand*\idlang[1]{\mathbf{1}_{#1}}
\newcommand*\tto[1]{\overset{#1}{\to}}

\newcommand{\mycomment}[1]{}

\begin{document}

\begin{frontmatter}

\title{Presheaf automata}

  \affiliation[gss]{organization={University of Sheffield},
              country={UK}}
  \affiliation[gsl]{organization={Collegium de Lyon},
              country={France}}
 \affiliation[kz]{organization={University of Warsaw},
              country={Poland}}
 \author[gss,gsl]{Georg Struth}
 \author[kz]{Krzysztof Ziemia\'nski}

\begin{abstract}
  We introduce presheaf automata as a generalisation of different
  variants of higher-dimensional automata and other automata-like
  formalisms, including Petri nets and vector addition systems. We
  develop the foundations of a language theory for them based on
  notions of paths and track objects. We also define open maps for
  presheaf automata, extending the standard notions of simulation and
  bisimulation for transition systems. Apart from these conceptual
  contributions, we show that certain finite-type presheaf automata
  subsume all Petri nets, generalising a previous result by van
  Glabbeek, which applies to higher-dimensional automata and safe
  Petri nets.
  We also present a class of presheaf automata for which there is no Kleene theorem with respect to the notions of rational and regular languages introduced.
\end{abstract}

%

\end{frontmatter}

\section{Introduction}

This work is motivated by the study of higher-dimensional automata
(HDA), a geometric model of concurrency introduced by Pratt and van
Glabbeek
\cite{Pratt_ModelingConcurrency,vanGlabbeek_Expressiveness}. Over
time, variants of the original formalisations have emerged
\cite{Goubault-Mimram_Relationship}, most recently, for instance, event
consistent HDA \cite{FJSZ_HDALang} and HDA with interfaces
\cite{FahrenbergJSZ24}. It therefore seems worth creating a formalism
in which these variants can be studied uniformly. Their common feature
is their formalisation as presheaves on similar, but different index
categories. The presheaf automata, which we introduce here, are
defined on a generalised index category and can not only be
instantiated to the variants of HDA mentioned, but also to other
automata-like models, including Petri nets and vector addition
systems.

A classical automaton over the alphabet $\Sigma$, for instance, can be
regarded as an edge-labelled directed graph (ignoring start and accept
states for now), formed by a set $V$ of vertices or states, a family
$(T_a)_{a\in\Sigma}$ of sets of $\Sigma$-labelled edges or
transitions, and of source and target maps $s_a,t_a:T_a\to V$.
Alternatively, this structure can be formalised as a presheaf on the
index category $\cG(\Sigma)$ with objects $\Sigma\cup \{\emptyset\}$
and (non-identity) morphisms $\sigma_a,\tau_a:\emptyset\to a$ for each
$a\in \Sigma$.  The elements of this presheaf are therefore vertices
and edges, its executions are sequences of elements in which every edge is
preceded by its source and succeeded by its target.

Classical automata can be viewed as one-dimensional HDA and HDA in
turn as higher-dimensional generalisations of classical automata. In
this context, vertices arise as $0$-cells, edges as $1$-cells, but
there may also be higher-dimensional cells, to which lower dimensional
cells can be attached by face maps, which are higher-dimensional
counterparts of the source and target maps of digraphs. When
formalised as presheafs, the index categories of HDA thus arise as
higher-dimensional generalisations of those of classical automata.

Here, we generalise this construction to $\cC$-automata para\-metrised
by a directed category or d-category $\cC$, which has distinguished
subcategories of ``source-like'' formorphisms $\cC^+$ and
``target-like'' backmorphisms $\cC^-$.  A $\cC$-automaton is then
essentially a presheaf on $\cC$ equipped with start and accept
elements, which is why we refer to $\cC$-automata less specifically as
``presheaf automata''. The generality of this new model stems from the
fact that d-categories provide a uniform way of modelling the index
categories of the HDA and automata-like models mentioned.  For- and
backmorphisms allow defining paths or executions of presheaf automata
by analogy to those of standard automata.

Among the presheaf automata, we distinguish a class of track objects
that are ``generated'' by paths.  These allow defining languages of
presheaf automata: they play the role of words of a language, and
accepting paths ``recognise'' the track objects they generate.
Further, maps from track objects to automata provide an alternative
notion of execution of automata, allowing track objects to be regarded
as ``path objects'' in the spirit of Joyal, Nielsen and
Winskel~\cite{Joyal-Winskel-Nielsen_Bisimulation}. This enables us to
define open maps and it leads to notions of (bi)simulations for presheaf
automata.  Finally, track objects can be compared using a subsumption
relation (a track is subsumed by another one if there is a morphism of
presheaf automata between them), generaling a similar notion from
\cite{FJSZ_HDALang}.
  
After introducing these fundamental concepts as the main contribution
of this article, we develop the foundations of a language theory for
presheaf automata, defining in particular classes of rational
languages, which can be presented by rational expressions, and regular
languages, which can be recognised by certain presheaf automata of
finite type.

In the final sections of this article we outline, as a proof of concept,
how known geometric models for concurrency, in particular HDA, can be
realised as presheaf automata on suitable index categories. We also
introduce a d-category $\cV_d$ such that $\cV_d$-automata are vector
addition systems with states, subsuming Petri nets.  Further, we show how to introduce counters to any d-category; in particular, we realise classical automata with one counter as a class of presheaf automata. It is well known that they accept non-rational languages, which rules out a general Kleene theorem for presheaf automata with respect to the notion of rational language introduced.
Finally, we
introduce higher-dimensional automata with counters (HDAC) as a special case of presheaf automata with counters.  Van
Glabbeek has shown that Petri nets can be realised by HDA
\cite{vanGlabbeek_Expressiveness}, but his result implies that
non-safe Petri nets lead to infinite HDA. As a technical contribution
of independent interest, we show that the safeness condition can be
lifted for HDACs, so that every Petri net can be realised as an HDAC
of finite type. For vector addition systems, presheaf automata with counters and HDAC, we also discuss their track objects and languages.

While our main emphasis lies on conceptual foundations and the link
between presheaf automata and models of concurrency, many
obvious questions about language-automata correspondences, topological
and geometric connections, or relationship to other kinds of automata
or other categorical approaches to automata remain unanswered. We
therefore conclude with some avenues for future work and a brief
discussion of its research context.

\section{Categories and Presheaves}
\label{s:categories-presheaves}

In this preliminary section we recall  basic facts about presheaves
from the
literature~\cite{MacLane_Categories,MacLane-Moerdijk_Sheaves,Kashiwara-Schapira_Categories},
as far as they are relevant to presheaf automata.

We restrict our attention to locally small categories, and start with
general notation for these.  We write $\Ob(\cC)$ for the objects of a
category $\cC$ and $\cC(U,V)$ for the set of morphisms from $U$ to $V$
in $\cC$.  We write $\simeq$ for isomorphisms and $\cong$ for natural
isomorphisms to emphasise the presence of functors and natural transformations.  As usual, $\Set$ and $\Cat$ denote the categories of
sets and of small categories.

Let $\cC$ be a small category.  \emph{A presheaf} on $\cC$ (a
\emph{$\cC$-presheaf}) is a functor $X:\cC^{\textup{op}}\to \Set$.  We
write $X[U]$ for the value of $X$ on the object $U$ of $\cC$ and
$X[\varphi]$ for the function $X[U]\to X[V]$ induced by the morphism
$\varphi:V\to U$ in $\cC$.  $\cC$-presheaves and their morphisms,
which are natural transformations, form the category $\psh\cC$.

Every object $U$ in $\cC$ defines the \emph{representable presheaf}
$\widehat{U}=\cC(-,U)$.  Hence $\widehat{U}[V] = \cC(V,U)$ for each
object $V$ in $\cC$ and
$\widehat{U}[\varphi]:\widehat{U}[V]\to \widehat{U}[W]$ for each
morphism $\varphi:W\to V$ in $\cC$, so 
$\widehat{U}[\varphi](\psi) = \psi\circ \varphi$ for each
$\psi:V\to U$.
A morphism $\alpha:U'\to U$
induces a natural transformation 
$\widehat{U}'\to\widehat{U}$	
with the components
$\cC(V,\alpha):\widehat{U}'[V]\to
\widehat{U}[V]$, 
$\cC(V,\alpha)(\psi) = \alpha\circ \psi$ for all
$\psi:V\to U'$ in $\cC$.
This defines the \emph{Yoneda embedding} functor
\begin{equation*}
	\yoneda_\cC:\cC\ni U\mapsto \widehat{U}\in \psh\cC,
\end{equation*}
which is full and faithful.
The Yoneda lemma guarantees that 
\begin{equation}
  \label{e:Yoneda}
  \psh\cC(\widehat{U},X)\ni f \mapsto f(\id_U)\in X[U]
\end{equation}
is a natural bijection for every $\cC$-presheaf $X$ and $U\in\Ob(\cC)$.

Each functor $F:\cD\to\cC$ induces the \emph{inverse image functor}
$F^*:\psh\cC\to\psh\cD$ defined by $F^*(X)[U]=X[F(U)]$, and its left
adjoint, the \emph{direct image functor} $F_*:\psh\cD\to\psh\cC$. As a
left adjoint, $F_\ast$ preserves colimits and representable
presheaves, that is, $F_*\widehat{V}\cong \widehat{F(V)}$ for every
$V\in\cD$.  

For a functor $F:\cD\to\cC$ and $U\in \Ob(\cC)$, we define the
category $F{\downarrow}U$ whose objects are pairs $(V,\phi)$ such that
$V\in \Ob(\cD)$ and $\phi:F(V)\to U$ and whose morphisms are given by
\begin{equation}
  (F{\downarrow}U)((W,\psi),(V,\phi))
  =\{\omega\in \cD(W,V)\mid \phi\circ F(\omega)=\psi\}.\label{e:SliceMorph}
\end{equation}
It is a special case of a comma category. We write
$\cD{\downarrow}U$ for the slice category $\Id_\cD{\downarrow}U$.

The \emph{category of elements} $\Cell{X}$ of a $\cC$-presheaf $X$ is
the category $\yoneda_{\cC}{\downarrow}X$.  It follows from
\eqref{e:Yoneda} that its objects -- the \emph{elements} of $X$ -- are
pairs $(U,x)$ such that $U\in\Ob(\cC)$ and $x\in X[U]$ while its
morphisms are given by
\begin{equation*}
  \Cell X ((V,y),(U,x))=\{(\phi,x)\mid \phi\in \cC(V,U),\; X[\phi](x)=y\}.
\end{equation*}

The category of elements is equipped with the forgetful functor
$\pi_X:\Cell X\to \cC$ (note that $X[U]\cong \pi_X^{-1}(U)$).  This
makes $\Cell$ into a functor $\psh\cC \to \Cat{\downarrow}\cC$ defined
by
\begin{equation}
  \label{e:CellFunctor}
  \Cell 
  X =
  (\Cell X ,\pi_X)
\end{equation}
on objects, while the morphisms in
  $\Cat{\downarrow}\cC$ are obtained from those in $\psh\cC$ according
  to \eqref{e:SliceMorph}.
It has a left adjoint described in Lemma~\ref{l:E-adj}
below. 

The $\cC$-presheaf \emph{generated} by a functor $G:\cE\to\cC$ from a
category $\cE$ is
\begin{equation}
\label{e:Gen}
  \widehat{G}=\colim(\yoneda_\cC\circ G)=\colim_{e\in\cE}
  \widehat{G(e)}.
\end{equation}

Informally, $G$ can be regarded as a presentation of the presheaf
$\widehat{G}$ by generators (values of $G$ on objects) and relations
(values on morphisms).  If $\cE\simeq\{*\}$ is trivial, then
$\widehat{G}\cong \widehat{G(*)}$.

The following \emph{density theorem}~\cite[Thm.\@
III.7.1]{MacLane_Categories} shows that every presheaf is generated by
the projection of its category of elements.
\begin{lem}
  \label{l:Density}
 If $X$ is a $\cC$-presheaf, then the family of maps
  \begin{equation*}
    \left(f_{(U,x)}:\widehat{\pi_X(U,x)}=\widehat{U}\xrightarrow{x}X\right)_{(U,x)\in\Cell{X}}
  \end{equation*}
  induces a natural isomorphism
  $\widehat{\pi_X}\xrightarrow\cong X$.
\end{lem}

Formula \eqref{e:Gen} defines a functor
$(\widehat{-}):\Cat{\downarrow}\cC\to\psh\cC$, because it is a
composition
\begin{equation}
\label{e:GenFun}
	(\widehat{-}):\Cat{\downarrow}\cC\xrightarrow{\Cat\slice\yoneda_\cC}
	\Cat\slice\psh\cC \xrightarrow{\colim} \psh\cC.
\end{equation}

\begin{lem}\label{l:E-adj}
  The functor $\widehat{(-)}:\Cat{\downarrow}\cC\to\psh\cC$
  is a left adjoint to $\Cell$.
\end{lem}

\begin{proof}
We show that for every $G:\cE\to\cC$ and $X\in\psh\cC$
there is a canonical isomorphism
\begin{align}
  \label{e:CellGenAdj}
  \psh\cC(\widehat{G},X)
  &\cong
  (\Cat{\downarrow}\cC)((\cE,G),(\Cell X,{\pi_X}))
  \\
  &=	
  \{F:\cE\to\Cell(X)\mid \pi_X\circ F=G\}.
\end{align}
Indeed, each functor $F\in (\Cat{\downarrow}\cC)((\cE,G),(\Cell X,\pi_X))$
induces a presheaf map $\widehat{G}\to \widehat\pi_X\cong X$.
Conversely, each presheaf map $f\in\psh\cC(\widehat{G},X)$
induces a functor $F:\cE\to\Cell X$ by
\[
  F(e)=(G(e),f(j_e(\id_{G(e)})))\qquad\text{ and }\qquad
  F(e'\xrightarrow\varepsilon e)
  =
  ( G(\varepsilon), f(j_e(\id_{G(e)}))),
\]
where $\id_{G(e)}\in \cE(G(e),G(e))=\widehat{G(e)}[G(e)]$
and $j_e:\widehat{G(e)}\to\widehat{G}$ is the structural map of the colimit.
\end{proof}

\begin{lem}
\label{l:Reflective}
  The functor $\Cell:\psh\cC\to\Cat{\downarrow}\cC$
  is full and faithful.
\end{lem}

\begin{proof}
  Lemma \ref{l:Density} and the adjunction in Lemma~\ref{l:E-adj}
  yield, for $X,Y\in\psh\cC$,
	\[
		\psh\cC(X,Y)
		\cong
		\psh\cC(\widehat{\pi_X},Y)
		\cong
		(\Cat{\downarrow}\cC)(\Cell X ,\Cell Y).\qedhere
	\]
\end{proof}
The image of $\Cell$ is thus a reflective subcategory of
$\Cat{\downarrow}\cC$, which we describe next.

A functor $F:\cD\to\cC$ is a \emph{discrete fibration} if for each
object $V$ of $\cD$ and morphism $\varphi:U\to F(V)$ of $\cC$ there is
a unique morphism $\psi:V'\to V$ in $\cD$ such that $F(\psi)=\varphi$.
Let $\Fib\subseteq \Cat$ denote the category of discrete fibrations
(compositions of discrete fibrations are again discrete fibrations).

For every $\cC$-presheaf $X$, $\pi_X:\Cell X\to \cC$ is a discrete
fibration: for each object $(U,x)$ in $\Cell{X}$ and morphism
$\varphi:V\to U$ in $\cC$, 
$(\varphi,x):(V,X[\varphi](x))\to (U,x)$ is the unique lifting of $\varphi$ with
target $(U,x)$.  Conversely, for every discrete fibration
$F:\cD\to\cC$, there is a $\cC$ presheaf $X$ defined by
$X[U]=F^{-1}(U)$ for each object $U$ in $\cC$ (the fibre of $U$ in
$\cD$) and for each morphism $\phi:U'\to U$ in $\cC$ by
$X[\phi]:F^{-1}(U)\to F^{-1}(U')$, such that for every
$V\in F^{-1}[U]$, $X[\phi](V)$ is the $V'\in F^{-1}(U')$ determined by
the unique lift $\psi:V'\to V$ of $\phi$.

These constructions lead to an equivalence of categories
\begin{equation}
\label{e:PShFib}
	\Cell:\psh\cC \xrightarrow\cong \Fib{\downarrow}\cC
\end{equation}
(see, for instance, \cite[Thm.\@ 2.1.2]{Loregian-Riehl_Categorical}
for a proof).  $\cC$-presheaves are therefore equivalent to discrete
fibrations over $\cC$, and
$\Fib{\downarrow}\cC\subseteq \Cat{\downarrow}\cC$ forms a reflective
subcategory, as illustrated in the diagram

\begin{equation}
\label{e:PresheafCategories}
\begin{tikzcd}
	\psh\cC \arrow[rr, "\Cell",shift left=1] \arrow[dr, "\cong",shift left=1] & & \Cat{\downarrow}\cC \arrow[ll, "\widehat{(-)}",shift left=1] \\
	& \Fib{\downarrow}\cC \arrow[ur,hook] \arrow[ul, "\cong",shift left=1]
\end{tikzcd}
\end{equation}
This summarises the relationship between presheaves,
categories of elements and discrete fibrations. 
The equivalence between presheaves and discrete
fibrations is used frequently in the sequel.

\section{Presheaf Automata}

In this section we introduce our main object of study. We start with
the index category for presheaf automata.

\begin{df}
  \label{DCat:Def}
  \emph{A directed category} (a \emph{d-category})
  is a triple $(\cC, \forM\cC, \bckM\cC)$
  where
  \begin{itemize}
  \item $\cC$ is a small category,
  \item $\forM\cC$ and $\bckM\cC$ are wide subcategories of
    \emph{forward morphisms} (\emph{formorphisms}) and \emph{backward morphisms} (\emph{backmorphisms}) of $\cC$,
    respectively,
      \item if a morphism $\varphi$ of $\cC$ is invertible, then
        $\varphi\in\forM\cC\iff \varphi^{-1}\in\bckM\cC$.
  \end{itemize}
\end{df}

We write $\fbM\cC=\forM\cC\cap \bckM\cC$, and $\isoM\cC$ for the
isomorphisms in $\fbM\cC$. We abbreviate $\cC^\alpha$ with
$\alpha\in\{+,-\}$. We call $(\forM\cC, \bckM\cC)$, or $\cC^\alpha$
for short, a \emph{d-structure} on $\cC$.

\begin{df}
  A \emph{d-functor} between d-categories $\cD$, $\cC$ is a functor
  $F:\cD\to \cC$ that preserves for- and backmorphisms:
  $F(\cD^\alpha)\subseteq \cC^\alpha$.  Small d-categories and d-functors
  form the category $\dCat$.
\end{df}

\begin{df}
  A \emph{natural d-transformation} between d-functors $F,G:\cD\to\cC$
  is a natural transformation $T:F\Rightarrow G$ such that
  $T(V)\in \fbM\cC(F(V),G(V))$ for all $V\in\Ob(\cD)$.
A \emph{d-equivalence} of d-categories is a
natural d-transformation that is an equivalence.
\end{df}

Equivalently, $F$ is a d-functor if it restricts to functors
$F^\alpha:\cD^\alpha\to\cC^\alpha$, and $T$ is a natural d-transformation
if it restricts to natural transformations
$T^\alpha:F^\alpha\Rightarrow G^\alpha$.

\begin{df}
\label{l:InducedDStructure}
Let $F:\cD\to\cC$ be a functor from a category $\cD$ to a d-category $\cC$.
The \emph{pullback} d-structure on $\cD$ 
is given by $\cD^\alpha=F^{-1}(\cC^\alpha)$.
\end{df}

For every $\cC$-presheaf $X$ on a d-category $\cC$, the category of elements $\Cell X$ is again a d-category with the pullback d-structure induced by $\pi_X:\Cell X\to\cC$.  The functor from \eqref{e:CellFunctor} thus lifts to $\Cell:\psh\cC \to \dCat{\downarrow}\cC$.

A d-functor $F:\cD\to\cC$ is a \emph{discrete d-fibration} if it is a
discrete fibration with the pullback d-structure on $\cD$ is induced
by $F$.  Clearly, $\pi_X$ is a discrete d-fibration.  We write
$\dFib\subseteq \dCat$ for the category of discrete d-fibrations.
Repeating the arguments from the previous section yields the following
fact.
\begin{prp}
\label{p:dAdj}
	For every d-category $\cC$,
	the functor $\widehat{(-)}:\dCat{\downarrow}\cC\to\psh\cC$
	is left adjoint to the functor $\Cell:\psh\cC\to \dCat{\downarrow}\cC$.
	Furthermore, $\Cell$ defines an equivalence of categories $\psh\cC\cong \dFib\slice\cC$.
      \end{prp}
      Thus \eqref{e:PresheafCategories} extends to the following
      diagram, where $\mathrm{ind}$ introduces the pullback d-structure from Definition
      \ref{l:InducedDStructure}.
\begin{equation}
\label{e:PresheafdCategories}
\begin{tikzcd}
	\psh\cC
		\arrow[rr, "\Cell",shift left=1]
		\arrow[dr, "\cong",shift left=1]
	& &
	\dCat{\downarrow}\cC
		\arrow[ll, "\widehat{(-)}",shift left=1]
		\arrow[rr, "\mathrm{forget}", shift left=1]
	& &
	\Cat\slice\cC
		\arrow[ll, "\mathrm{ind}",shift left=1]
\\
	& \dFib{\downarrow}\cC
		\arrow[ur,hook]
		\arrow[ul, "\cong",shift left=1]
		\arrow[rr, "\cong",shift left=1]
	&&
	\Fib\slice\cC
		\arrow[ur,hook]
		\arrow[ll, "\cong",shift left=1]
\end{tikzcd}
\end{equation}

We are now prepared for our most important definition.
\begin{df}
  \label{d:Aut}
  A \emph{presheaf automaton} on a d-category $\cC$ (a
  \emph{$\cC$-automaton}) is a $\cC$-presheaf $X$ with distinguished
  sets of \emph{start elements} $\bot_X$ and \emph{accept elements}
  $\top_X$ in $\Ob(\Cell X)$.
\end{df}

The category $\aut\cC$ has $\cC$-automata as objects. Its morphisms
are morphisms of $\cC$-presheaves that preserve start and accept
elements:
\begin{equation}
  \aut\cC(Y,X)=\{f\in\psh\cC(Y,X)\mid f(\bot_Y)\subseteq \bot_X,\; f(\top_Y)\subseteq \top_X\}.
\end{equation}
Because of \eqref{e:PresheafdCategories}, 
$\cC$-automata are equivalent to discrete d-fibrations over $\cC$
with distinguished start and accept objects.

Initial examples of presheaf automata are discussed in Section~\ref{s:StdAut} and \ref{s:PrecubicalSets}; for more advanced ones see Section~\ref{s:HDA} - \ref{s:HDAC}.

\begin{df}
\label{d:SimpleAutomata}
A $\cC$-automaton $X$ is \emph{simple} if it has one single start and
one single accept element, denoted (with some abuse of notation) by
$\bot_X$ and $\top_X$, respectively.  The \emph{source} of a simple
$\cC$-automaton $X$ is defined as $\src(X)=\pi_X(\bot_X)\in\Ob(\cC)$,
its \emph{target} as $\tgt(X)=\pi_X(\top_X)\in\Ob(\cC)$.
\end{df}

The Yoneda lemma allows us to interpret start and accept elements as
maps from representable presheaves.  A simple automaton $X$ can thus
be regarded as a cospan of presheaves
\begin{equation}
\label{e:SimpleAutSpan}
\widehat{\src(X)}\xrightarrow{\bot_X} X \xleftarrow{\top_X} \widehat{\tgt(X)}.
\end{equation}
If $f:X\to Y$ is a morphism of simple $\cC$-automata, then
\begin{equation*}
  U:=\src(X)=\src(Y)\qquad\text{ and }\qquad V:=\tgt(X)=\tgt(Y).
\end{equation*}
In terms of cospans,
$f$ is represented by the following commutative diagram in $\psh\cC$:
\begin{equation}
		\begin{tikzcd}[row sep=tiny,column sep=large]
			 & X\arrow{dd}[left]{f} & \\
			\widehat{U}\arrow{ur}[above left]{\bot_X} \arrow{dr}[below left]{\bot_Y} &  &
			\widehat{V}\arrow{ul}[above right]{\top_X} \arrow{dl}[below right]{\top_Y} \\
			 & Y & 
		\end{tikzcd}
\end{equation}
Simple automata form a full subcategory $\saut\cC\subseteq \aut\cC$,
which decomposes into a disjoint union
\begin{equation}
  \saut\cC=\coprod_{U,V\in\Ob(\cC)}\ilo{U}{\saut\cC}{V},
\end{equation}
where $\ilo{U}{\saut\cC}{V}$ is the full subcategory of $\saut\cC$ with objects
$X$ satisfying $\src(X)=U$ and $\tgt(X)=V$.

\section{Standard automata as presheaf automata}
\label{s:StdAut}

In this and the following section, we present two examples of d-categories and illustrate some of the concepts and constructions from the previous section. Here, we show that presheaf automata on an appropriate d-category closely correspond to standard automata. In the next section, we do the same for precubical sets, that is, higher dimensional automata without labels.

As usual for automata, we fix an alphabet $\Sigma$.  An index d-category
$\cG=\cG(\Sigma)$ for standard automata is given by the following data:
	\begin{itemize}
        \item $\Ob(\cG)=\{\vo\}\cup \Sigma$ for $\vo\not\in\Sigma$,
  \item $\cG(x,x)=\{\id_x\}$ for $x\in \Ob(\cG)$,
  \item $\cG(\vo,a)=\{\sigma_a,\tau_a\}$
  and  $\cG(a,\vo)=\emptyset$ for $a\in\Sigma$,
  \item $\forM{\cG}=\{\sigma_a\}_{a\in\Sigma}\cup \{\id_x\}_{x\in \Ob(\cG)}$,
    $\bckM{\cG}=\{\tau_a\}_{a\in\Sigma}\cup \{\id_x\}_{x\in \Ob(\cG)}$.	
  \end{itemize}

\begin{exa}
\label{x:GdCat}
For $\Sigma=\{a, b\}$ and omitting identity morphisms, we may
illustrate $\cG$ as
\begin{equation*}
  \begin{tikzcd}[column sep = large]
    a & \vo\ar[l,shift left, "\tau_a"]\ar[l, shift right, swap,
    "\sigma_a"]\ar[r, shift left, "\sigma_b"]\ar[r, shift right, swap,
    "\tau_b"] & b
  \end{tikzcd}.
\end{equation*}
\end{exa}

Next, we translate standard automata to presheaf automata
on $\cG$.  To any standard automaton, defined as a labelled digraph $(Q,T,\lambda,\src,\tgt,S,F)$ of a set $Q$ of states, a set $T$ of transitions, a labelling function $\lambda:T\to\Sigma$, source and target maps $\src,\tgt:T\to Q$, as indicated in the introduction, and sets $S,F\subseteq Q$ of start and accept states, respectively, we assign a $\cG$-automaton $X$ such that 
\begin{itemize}
\item $X[\vo]=Q$ is the set of vertices,
\item $X[a]=\lambda^{-1}(a)$ is the set of $a$-labelled edges,
\item $X[\sigma_a]=\src|_{\lambda^{-1}(a)}$
	and
	$X[\tau_a]=\tgt|_{\lambda^{-1}(a)}$
	are the restrictions of the source and target maps, respectively,
\item $\bot_X=\{(\vo,v)\mid v\in S\}$ and $\top_X=\{(\vo,v)\mid v\in F\}$.
\end{itemize}

\begin{exa}
  To the standard automaton
\begin{equation}
 \label{e:StdAutExample}
 \vcenter{\hbox{%
 \begin{tikzpicture}[node distance = 2cm]
   \node[state,initial] (x) { $x$};
   \node[state,accepting] (y) [right=of x]{$y$};
   \path (x) edge[bend left] node {$a$} (y);
   \path (y) edge[bend left] node {$b$} (x);
   \path (y) edge[loop, out = 335, in = 25, looseness = 20] node[swap]
   {$b$} (y);
 \end{tikzpicture}
 }}
\end{equation}
 we associate the
  $\cG$-presheaf $X$ with $X[\vo]=\{x,y\}$, $X[a]=\{e_1\}$ and
  $X[b]=\{e_2,e_3\}$ on objects of $\cG$ and
  $X[\sigma_a]:e_1\mapsto x$, $X[\sigma_b]:e_2\mapsto y,e_3\mapsto y$,
  $X[\tau_a]:e_1\mapsto y$ and $X[\tau_b]:e_2\mapsto x,e_3\mapsto y$
  on morphisms ($e_1$, $e_2$ and $e_3$ are just names of transitions). Its category of elements $\Cell{X}$ and the projection
  functor $\pi_X:\Cell{X}\to\cG$ are shown in the following diagram, where
  identity arrows have been omitted and $\pi_X$ is indicated by dashed lines:
 \begin{equation}
 \label{e:GFibExample}
   \begin{tikzcd}
     & (a,e_1)\ar[rrrrr,dashed,very thin]&&&&&a\\
     (\vo,x)\ar[ur, "{(\sigma_a,e_1)}"]\ar[dr, swap,
     "{(\tau_b,e_2)}"]\ar[rrrrrr,dashed,very thin,bend left=5]&& (\vo,y)\ar[ul, swap, "{(\tau_a,e_1)}"]\ar[dl,"{(\sigma_b,e_2)}"]\ar[drr,
     shift left, "{(\sigma_b,e_3)}"]\ar[drr, shift right, swap,
     "{(\tau_b,e_3)}"]
     \ar[rrrr,dashed,very thin] &&&&\vo\ar[u,shift left, "\tau_a"]\ar[u, shift right, swap,
    "\sigma_a"]\ar[d, shift left, "\sigma_b"]\ar[d, shift right, swap,
    "\tau_b"] \\
     &(b,e_2)\ar[rrrrr,dashed,very thin,bend right=9]&&&(b,e_3)\ar[rr,dashed,very thin]&&b
   \end{tikzcd}
 \end{equation}
 Assigning $\bot_X=\{(\vo,x)\}$ and $\top_X=\{(\vo,y)\}$
 makes $X$ a $\cG$-automaton.
Obviously, $\pi_X$ is the discrete fibration associated with $X$ by \eqref{e:PShFib}.
\end{exa}

Conversely, from every $\cG$-automaton $X$ we recover a standard
automaton $(Q,T,\lambda,\src,\tgt,S,F)$ given by
\begin{gather*}
  Q=X[\vo], \qquad
  T=\coprod_{a\in\Sigma} X[a], \qquad
  \lambda=\pi_X|_{\coprod_{a\in\Sigma} X[a]}, \\
  \src=\coprod_{a\in\Sigma} X[\sigma_a], \qquad
  \tgt=\coprod_{a\in\Sigma} X[\tau_a],\\
  S=\{x\in X[\vo] \mid (\vo,x)\in \bot_X\}, \qquad
  T=\{x\in X[\vo] \mid (\vo,x)\in \top_X\}.
\end{gather*}
The projection $\pi_X$ indicates whether an object of $\Cell{X}$
represents a vertex or an edge with its label.  The labels of the
automaton \eqref{e:StdAutExample} are thus obtained by projecting onto
the index category in \eqref{e:GFibExample}.

Formally, our definition of standard automata is slightly more general
than the textbook one, as it allows parallel edges with the same label
and start and accept transitions. As this difference does not
affect expressivity, we consider both as essentially the same.

\section{Precubical sets}
\label{s:PrecubicalSets}

As a second example, we now model the precubical category and
precubical sets using presheaf automata.  These form a basis for defining
higher-dimensional automata, geometrical models of
concurrency~\cite{FahrenbergJSZ24, Goubault-Mimram_Relationship,
  Grandis-Mauri_Cubical, vanGlabbeek_Expressiveness}, as instances of
presheaf automata in Section \ref{s:HDA}. Cubical sets were introduced
to topology by Serre and Kan~\cite{Kan55,Serre51}. Here we present a standard definition of a precubical category~\cite{Antolini00}, see
also~\cite{Crans95,Grandis-Mauri_Cubical,Jardine02} for the cubical case.

\begin{df}
\label{d:PrecubeCategory}
  The \emph{precubical category} $\oldsq$
  is the following d-category:
  \begin{itemize}
  \item
    objects are symbols $[n]$ for $n\geq 0$,
  \item morphisms are generated by
    $d^0_{i}, d^1_{i}\in\oldsq([n-1],[n])$ for $1\leq i\leq n$ and the
    relations
    $d^\varepsilon_{j}\circ d^\eta_{i}=d^\eta_{i}\circ
    d^\varepsilon_{j-1}$, for $1\leq j<i\leq n$,
    $\varepsilon,\eta\in\{0,1\}$,
  \item
    $\forM\oldsq$ is generated by $\{d^0_i:[n-1]\to[n]\}$ for $1\leq i\leq n$,
  \item
    $\bckM\oldsq$ is generated by $\{d^1_i:[n-1]\to[n]\}$ for $1\leq i\leq n$.
  \end{itemize}
\end{df}

For $m\leq n$, $\varepsilon\in\{0,1\}$ and
$A=\{a_1<a_2<\dotsm<a_{n-m}\}\subseteq\{1,\dotsc,n\}$, we define
\begin{equation}
		d^\varepsilon_A=d^\varepsilon_{a_n}\circ d^\varepsilon_{a_{n-1}}\circ\dotsm
		d^\varepsilon_{a_2}\circ d^\varepsilon_{a_{1}}.
\end{equation}
For- and backmorphisms in $\oldsq$ are the morphisms $d^0_A$ and $d^1_A$, for
$A\subseteq\{1,\dotsc,n\}$, respectively.

A \emph{precubical set} in the sense of \cite[Def.\@
1]{vanGlabbeek_Expressiveness} and \cite[Def.\@ 6.1]{FGHMR_DAT} is a
$\oldsq$-presheaf $X$: a family of cells $(X[n])_{n\geq 0}$ of
dimension $n$ with (elementary) \emph{face maps}
$\delta^\varepsilon_i=X[d^\varepsilon_i]:X[n]\to X[n-1]$, which satisfy
the cubical identities
$\delta^\varepsilon_i\circ \delta^\eta_j=\delta^\eta_{j-1}\circ
\delta^\varepsilon_i$ for $i<j$ and $\varepsilon,\eta\in\{0,1\}$
that are imposed by the relations between the morphisms of
$\oldsq$.

\begin{exa}
\label{x:oldsq1}
An example of a $\oldsq$-automaton $X$ in the sense of Definition~\ref{d:Aut}, and hence of a precubical
  set, is shown below:
  
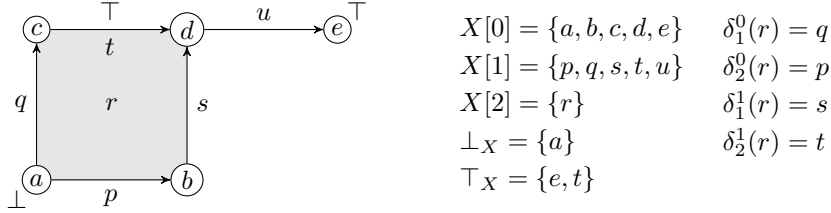
\[
    \begin{tikzpicture}
      \filldraw[color=black!10!white] (0,0)--(2,0)--(2,2)--(0,2)--(0,0);
      \node[state, minimum size=10pt] (00) at (0,0) {$a$};
      \node[state, minimum size=10pt] (10) at (2,0) {$b$};
      \node[state, minimum size=10pt] (01) at (0,2) {$c$};
      \node[state, minimum size=10pt] (11) at (2,2) {$d$};
      \node[state, minimum size=10pt] (21) at (4,2) {$e$};
      \path (00) edge node[below] {$p$} (10);
      \path (01) edge node[above] {$\top$} node[below] {$t$} (11);
      \path (11) edge node[above] {$u$} (21);
      \path (00) edge node[left] {$q$} (01);
      \path (10) edge node[right] {$s$} (11);
      \node at (1,1) {$r$};
      \node[below left] at (00) {$\bot$};
      \node[above right] at (21) {$\top$};
      \node[right] at (5.5,2) {$X[0]=\{a,b,c,d,e\}$};
      \node[right] at (5.5,1.5) {$X[1]=\{p,q,s,t,u\}$};
      \node[right] at (5.5,1) {$X[2]=\{r\}$};
      \node[right] at (5.5,0.5) {$\bot_X=\{a\}$};
      \node[right] at (5.5,0) {$\top_X=\{e,t\}$};
      \node[right] at (9,2) {$\delta^0_1(r)=q$};
      \node[right] at (9,1.5) {$\delta^0_2(r)=p$};
      \node[right] at (9,1) {$\delta^1_1(r)=s$};
      \node[right] at (9,0.5) {$\delta^1_2(r)=t$};
    \end{tikzpicture}
  \]
  The data on the right defines $X$. The picture on the left is a visualisation, and in fact a geometric realisation of $X$ in the
  sense of \cite{FGHMR_DAT}.  Each edge $x\in X[1]$ has
  source $\delta^0_1(x)$ and target $\delta^1_1(x)$. For instance,
  $\delta^0_1(s)=b$.
\end{exa}

$\oldsq$-automata are closely related to the higher-dimensional
automata of \cite{FJSZ_HDALang, FahrenbergJSZ24,
  vanGlabbeek_Expressiveness}.  We do not introduce the usual labels
on edges and higher cells at this stage as we want to focus on
the underlying cubical structure and keep things simple. Various modifications of the precubical category are
discussed in Sections~\ref{s:HDA} and \ref{s:HDAC} below.

\section{Paths}
\label{s:paths}

Next, we define executions of presheaf automata as paths, by analogy
to the paths generated by digraphs. It is convenient to start with an
intuitive notion, using d-categories as index categories, where paths
in presheaves appear as paths in their categories of elements.  We
then extend this definition and reformulate it
functorially. Throughout this section, $\cC$ is a d-category.

\begin{df}
\label{d:PathClassical}
A \emph{path} in a d-category $\cC$ is a sequence
\begin{equation}
\label{e:Path}
  \omega=(x_0,\phi_1,x_1,\phi_2,\dotsc,x_{n-1},\phi_n,x_n),
\end{equation}
where $x_k\in\Ob(\cC)$ and either $\phi_k\in \forM\cC(x_{k-1},x_k)$ or
$\phi_k\in \bckM\cC(x_{k},x_{k-1})$ for every $1 \le k\le n$.  In the
first case, we call $(x_{k-1},\phi_k ,x_k)$ an \emph{upstep} and write
\hbox{$x_{k-1} \arrO{\phi_k} x_k$} (with or without parentheses); in the
second, we speak of a \emph{downstep} and write
$x_{k-1}\arrI{\phi_k} x_k$.  Collectively, up- and downsteps are
called \emph{steps}.

The \emph{source} and \emph{target} of the path $\omega$ are
$\src(\omega)=x_0$ and $\tgt(\omega)=x_n$, respectively.  The \emph{concatenation} of paths $\omega=(x_0,\varphi_1,\dotsc,x_n)$ and
$\eta=(y_0,\psi_1,\dotsc,y_m)$ is defined as
\[
	\omega*\eta = (x_0,\varphi_1,\dotsc,\varphi_n,x_n,\psi_1,\dotsc,\psi_m,y_m)
      \]
      whenever $x_n=y_0$.  Every path is thus a concatenation of steps.
\end{df}

\begin{df}
  A \emph{path} in a $\cC$-presheaf $X$ is a path in its d-category of
  elements $\Cell X$.  A path $\alpha$ in a $\cC$-automaton $X$ is
  \emph{accepting} if $\src(\alpha)\in\bot_X$ and
  $\tgt(\alpha)\in\top_X$.
\end{df}

\begin{exa}
  Let $X$ be a $\cG$-presheaf, as defined in Section~\ref{s:StdAut}.
  Each path in $X$ that starts and terminates in a vertex or state is a
  sequence
 \[
 	\big(v_0 \arrO{\sigma_{a_1}}
 	e_1  \arrI{\tau_{a_1}} 
 	v_1 \arrO{\sigma_{a_2}}
 	e_2\arrI{\tau_{a_2}}
 	\dotsm
 	\arrO{\sigma_{a_{n}}}
 	e_{n} \arrI{\tau_{a_{n}}}
 	v_{n}\big)
 \]
 where $v_{k}\in X[\vo]$, $a_{k}\in\Sigma$ and
 $e_{k}\in X[a_{k}]$ for all $k$.  Up- and downsteps obviously
 alternate. They may be interspersed with identities, which are both
 up- and downsteps. Each upstep $\sigma_{a_k}$ starts an $a_k$-labeled
 transition $e_k$ while the following downstep $\tau_{a_k}$ terminates
 this transition.  
 Paths can thus be viewed as alternating sequences of states and transitions, as usual.
\end{exa}

Next, we develop an equivalent more categorical notion.
Let $\mathbf{1}$ denote the trivial d-category with one object
and the identity morphism only.

\begin{df}
  A \emph{bipointed d-category} is a d-category $\cC$ with two
  distinguished objects, the \emph{basepoints} $\bot_\cC$ and
  $\top_\cC$.  A \emph{bipointed functor} $F:\cD\to\cC$ preserves
  basepoints: $F(\bot_\cD)=\bot_\cC$ and
  $F(\top_\cD)=\top_\cC$.  The \emph{concatenation} of bipointed
  d-categories $\cC$ and $\cD$ is the bipointed d-category
  \[
    \cC*\cD=\colim \left( \cC \xleftarrow{\top_\cC} \bI \xrightarrow{\bot_\cD} \cD  \right).
  \]
  Also, $\bot_{\cC*\cD}$ and $\top_{\cC*\cD}$ are the images of
  $\bot_\cC$ and $\top_\cD$ under suitable colimit injections.
\end{df}

The concatenation $*$ is associative and
functorial: a pair of bipointed d-functors $\cC\to\cC'$, $\cD\to\cD'$
induces a bipointed d-functor $\cC*\cD\to\cC'*\cD'$.
This equips bipointed d-categories with a monoidal structure,
which is restricted from the cospan category of d-categories.

We define the  d-categories
\begin{equation}
\label{e:ElementaryLinear}
  \upstep
  =
  \big(\!
    \begin{tikzcd}	 	
      \bot \arrow[r,"\sigma"] & \top
    \end{tikzcd}
  \!\big),
  \qquad
  \downstep
  =
  \big(\!
    \begin{tikzcd}	 	
      \bot & \arrow[l,"\tau"] \top
    \end{tikzcd}
  \!\big),
  \qquad	 
  \isostep
  =
  \big(\!
    \begin{tikzcd}	 	
      \bot \arrow[r, shift left,"\sigma"] & \arrow[l,shift left, "\tau"] \top
    \end{tikzcd}
    \!\big).
\end{equation}
The morphisms $\sigma$ are formorphisms,
$\tau$ are backmorphisms,
and $\sigma=\tau^{-1}$ in $\isostep$.

\begin{df}
\label{d:LinearCategory}
A bipointed d-category $\cI$ is \emph{linear} if it is isomorphic to a
concatenation of elementary linear categories from
\eqref{e:ElementaryLinear}, and hence of the form
$\cI_1*\dotsm*\cI_n$ for $n\ge 0$ and
$\cI_k\in\{\upstep, \downstep, \isostep\}$.
\end{df}
Linear categories with basepoint-preserving d-functors as morphisms
form the category $\Lin$.

\begin{df}
  \label{d:Path}
  A \emph{path} in $\cC$ is a d-functor $\omega:\cI\to\cC$ from a
  linear category $\cI$.  A \emph{path} in a $\cC$-presheaf $X$ is a
  path in $\Cell{X}$.  The \emph{source} of $\omega$ is
  $\src(\omega)=\omega(\bot_\cI)$; its \emph{target} is
  $\tgt(\omega)=\omega(\top_\cI)$. The \emph{concatenation} of paths $\omega:\cI\to\cC$ and
  $\psi:\cJ\to\cC$ with $\tgt(\omega)=\src(\psi)$ is the path
  $\omega*\psi:\cI*\cJ\to \cC$ owing to the universal property of
  colimits. 
\end{df}
For any d-category $\cC$ there are natural bijections
\begin{equation*}
	\dCat(\upstep,\cC)\cong \forM\cC,\qquad
 	\dCat(\downstep,\cC)\cong \bckM\cC,\qquad
        \dCat(\isostep,\cC)\cong \isoM\cC.
 \end{equation*}
 A path $\omega:\upstep\to \cC$ is thus an upstep
 $\big(\omega(\bot) \arrO{\omega(\sigma)}\omega(\top)\big)$ and a path
 $\omega:\downstep\to \cC$ a downstep
 $\big(\omega(\bot) \arrI{\omega(\tau)}\omega(\top)\big)$, while a path
 $\omega:\isostep\to\cC$ may be regarded as both. Once again, such
 paths are called \emph{steps}. The following fact is obvious.
\begin{lem}
\label{l:PathConcatSteps}
	Every path is isomorphic to a concatenation of steps.
\end{lem}

A path \eqref{e:Path} in $\cC$ the sense of Definition \ref{d:PathClassical}
is interpreted as a functor
$\cI=\cI_1*\dotsc*\cI_n\to \cC$,
where $\cI_k=\upstep$ if $\varphi_k$ is a formorphism
and $\cI_k=\downstep$ if $\varphi_k$  a backmorphism.
For example, a path
\begin{equation*}
 \big(x_0 \arrO{\varphi_1}
  x_1 \arrI{\varphi_2} 
  x_2 \arrI{\varphi_3} 
  x_3 \arrO{\varphi_4}
  x_4\big)
\end{equation*}
relates to the functor $\upstep*\downstep*\downstep*\upstep\to\cC$
that sends the generating morphisms of factors to $\varphi_1$,
$\varphi_2$, $\varphi_3$, $\varphi_4$.

Paths in $\cC$ form the category
$\Path(\cC)=\Lin\slice\cC$: objects are paths $\omega:\cI\to\cC$, while morphisms from $\omega$ to $\psi:\cJ\to\cC$ are functors $F:\cI\to\cJ$ which preserve basepoints and satisfy $\omega=\psi\circ F$.

\begin{df}
  \label{d:PathRefinement}
  We define a relation $\denser$ on the set of paths where
  $\psi\denser\omega$ if there is a morphism $\omega\to \psi$ in
  $\Path(\cC)$. The paths $\omega$ and $\psi$ are \emph{equivalent},
  $\omega\simeq\psi$, if they lie in the same connected component of
  $\Path(\cC)$.
\end{df}

Equivalent paths must have the same sources and targets.  The relation
$\denser$ is a preorder and $\simeq$ is the least equivalence
containing $\denser$.  In fact, $\preceq$ is a precongruence with
respect to $\ast$, that is, $\omega_1\denser \psi_1$ and
$\omega_2\denser \psi_2$ imply
\hbox{$\omega_1*\omega_2\denser \psi_1*\psi_2$}, and $\simeq$ is a congruence
with respect to $\ast$.

The following two lemmas describe $\denser$ more explicitly.  The
first follows by an elementary calculation, the second is its
straightforward consequence.
\begin{lem}
\label{l:ElemPathEqs}
Basepoint-preserving d-functors between linear categories are
generated by concatenations of the basepoint-preserving d-functors
\[
\upstep\to \upstep*\upstep, \quad
\downstep\to \downstep*\downstep,\quad
\upstep\to\isostep,\quad
\downstep\to\isostep,\quad
\isostep\to \bI.
\]
All these d-functors are uniquely determined by their domains and
codomains.
\end{lem}

\begin{lem}
  \label{l:ClRefin}
  The precongruence $\denser$ is generated by the elementary relations
  \begin{enumerate}
  \item
    $(x\arrO{\phi}y\arrO{\psi}z)\denser (x\arrO{\psi\circ \phi}z)$ and 
    $(x\arrI{\phi}y\arrI{\psi}z)\denser (x\arrI{\psi\circ \phi}z)$,
  \item $(x)\denser(x\arrO{\id}x)$ and $(x)\denser(x\arrI{\id}x)$,
  \item if $\phi=\psi^{-1}$, then
    $(x\arrO{\phi}y)\denser (x\arrI{\psi}y)$ and $(x\arrI{\psi}y)\denser (x\arrO{\phi}y)$,
  \item if $\omega\denser \omega'$, then $\eta*\omega*\eta'\denser \eta*\omega'*\eta'$.
  \end{enumerate}
  The congruence $\simeq$ is generated by the same relations with
  $\denser$ replaced by $\simeq$.
\end{lem}

\begin{exa}
\label{x:PathsInG}
  In paths of $\cG$-automata, up- and downsteps alternate, except for
  identities, since non-identity formorphisms cannot be composed;
  and likewise for backmorphisms.  The
  elementary relations in Lemma~\ref{l:ClRefin} thus reduce to
    $(x)\denser(x\arrO{\id}x)$ and $(x)\denser(x\arrI{\id}x)$.
    Every path in $\cG$ is therefore equivalent to a sequence
    \[
\sigma_{a_1} \tau_{a_1} \sigma_{a_2} \tau_{a_2} \dotsm \sigma_{a_n} \tau_{a_n},
    \]
    possibly preceded by $\tau_{a_0}$ and succeeded by $\sigma_{a_{n+1}}$, for $a_k\in\Sigma$.
\end{exa}

\begin{exa}
  For paths in $\oldsq$ (Section \ref{s:PrecubicalSets}),
  \begin{equation*}
    ([0]\arrO{d^0_1}[1]\arrO{d^0_2}[2])
    \preceq
    ([0]\arrO{d^0_{1,2}}[2])
    \succeq
    ([0]\arrO{d^0_1}[1]\arrO{d^0_1}[2])
    .
  \end{equation*}
  All three paths are equivalent.
\end{exa}

\begin{exa}
  \label{ex:hdapaths}
  Let $X$ be the $\oldsq$-presheaf on the left below:
    \[
      \begin{tikzpicture}[scale=0.7]
	\filldraw[color=black!10!white] (0,0)--(2,0)--(2,2)--(0,2)--(0,0);
  \node[state, minimum size=10pt] (00) at (0,0) {$a$};
  \node[state, minimum size=10pt] (10) at (2,0) {$b$};
  \node[state, minimum size=10pt] (01) at (0,2) {$c$};
  \node[state, minimum size=10pt] (11) at (2,2) {$d$};
  \path (00) edge node[below] {$p$} (10);
  \path (01) edge node[above] {$r$} (11);
  \path (00) edge node[left] {$q$} (01);
  \path (10) edge node[right] {$s$} (11);
  \node at (1,1) {$u$};
   \begin{scope}[shift={(3.5,0)}]
	\filldraw[color=black!10!white] (0,0)--(2,0)--(2,2)--(0,2)--(0,0);
  \node[state, minimum size=10pt] (00) at (0,0) {};
  \node[state, minimum size=10pt] (10) at (2,0) {};
  \node[state, minimum size=10pt] (01) at (0,2) {};
  \node[state, minimum size=10pt] (11) at (2,2) {};
  \path (00) edge (10);
  \path (01) edge (11);
  \path (00) edge (01);
  \path (10) edge (11);
  \node[circle, inner sep=2pt, fill=blue] (a00) at (0,0) {};
  \node[circle, inner sep=2pt, fill=blue] (a10) at (1,0) {};
  \node[circle, inner sep=2pt, fill=blue] (a11) at (1,1) {};
  \node[circle, inner sep=2pt, fill=blue] (a22) at (2,2) {};
  \path (a00) edge[blue,thick] (a10);
  \path (a10) edge[blue,thick] (a11);
  \path (a11) edge[blue,thick] (a22);
  \node[above left,blue] at (a11) {$\alpha$};
	\end{scope}    
   \begin{scope}[shift={(7,0)}]
	\filldraw[color=black!10!white] (0,0)--(2,0)--(2,2)--(0,2)--(0,0);
  \node[state, minimum size=10pt] (00) at (0,0) {};
  \node[state, minimum size=10pt] (10) at (2,0) {};
  \node[state, minimum size=10pt] (01) at (0,2) {};
  \node[state, minimum size=10pt] (11) at (2,2) {};
  \path (00) edge (10);
  \path (01) edge (11);
  \path (00) edge (01);
  \path (10) edge (11);
  \node[circle, inner sep=2pt, fill=blue] (a00) at (0,0) {};
  \node[circle, inner sep=2pt, fill=blue] (a11) at (1,1) {};
  \node[circle, inner sep=2pt, fill=blue] (a22) at (2,2) {};
  \path (a00) edge[blue,thick] (a11);
  \path (a11) edge[blue,thick] (a22);
  \node[above left,blue] at (a11) {$\beta$};
	\end{scope}    
   \begin{scope}[shift={(10.5,0)}]
	\filldraw[color=black!10!white] (0,0)--(2,0)--(2,2)--(0,2)--(0,0);
  \node[state, minimum size=10pt] (00) at (0,0) {};
  \node[state, minimum size=10pt] (10) at (2,0) {};
  \node[state, minimum size=10pt] (01) at (0,2) {};
  \node[state, minimum size=10pt] (11) at (2,2) {};
  \path (00) edge (10);
  \path (01) edge (11);
  \path (00) edge (01);
  \path (10) edge (11);
  \node[circle, inner sep=2pt, fill=blue] (a00) at (0,0) {};
  \node[circle, inner sep=2pt, fill=blue] (a01) at (0,1) {};
  \node[circle, inner sep=2pt, fill=blue] (a11) at (1,1) {};
  \node[circle, inner sep=2pt, fill=blue] (a21) at (2,1) {};
  \node[circle, inner sep=2pt, fill=blue] (a22) at (2,2) {};
  \path (a00) edge[blue,thick] (a01);
  \path (a01) edge[blue,thick] (a11);
  \path (a11) edge[blue,thick] (a21);
  \path (a21) edge[blue,thick] (a22);
  \node[above left,blue] at (a11) {$\gamma$};
	\end{scope}    
   \begin{scope}[shift={(14,0)}]
	\filldraw[color=black!10!white] (0,0)--(2,0)--(2,2)--(0,2)--(0,0);
  \node[state, minimum size=10pt] (00) at (0,0) {};
  \node[state, minimum size=10pt] (10) at (2,0) {};
  \node[state, minimum size=10pt] (01) at (0,2) {};
  \node[state, minimum size=10pt] (11) at (2,2) {};
  \path (00) edge (10);
  \path (01) edge (11);
  \path (00) edge (01);
  \path (10) edge (11);
  \node[circle, inner sep=2pt, fill=blue] (a00) at (0,0) {};
  \node[circle, inner sep=2pt, fill=blue] (a10) at (1,0) {};
  \node[circle, inner sep=2pt, fill=blue] (a20) at (2,0) {};
  \node[circle, inner sep=2pt, fill=blue] (a21) at (2,1) {};
  \node[circle, inner sep=2pt, fill=blue] (a22) at (2,2) {};
  \path (a00) edge[blue,thick] (a10);
  \path (a10) edge[blue,thick] (a20);
  \path (a20) edge[blue,thick] (a21);
  \path (a21) edge[blue,thick] (a22);
  \node[above left,blue] at (a10) {$\zeta$};
	\end{scope}    
    \end{tikzpicture}
	\]
        Then $\beta=(a\arrO{d^0_{1,2}}u\arrI{d^1_{1,2}}d)$ is
        $\preceq$-smaller than both
        $\alpha=(a\arrO{d^0_1}p\arrO{d^0_2}u\arrI{d^1_{1,2}}d)$ and
        $\gamma=(a \arrO{d^0_1} q \arrO{d^0_1} u \arrI{d^1_1} s
        \arrI{d^1_1}d)$.  Thus, $\alpha\simeq \gamma$ but neither
        $\alpha\denser\gamma$ nor $\gamma\denser\alpha$.  The path
        $\zeta=(a \arrO{d^0_1} p \arrI{d^1_1} b \arrO{d^1_1} s
        \arrI{d^1_1}d)$ is not equivalent to the others.
\end{exa}

We do not aim to take the category of (single) pointed
d-categories as a category of models in the sense of
\cite{Joyal-Winskel-Nielsen_Bisimulation}, and the category of linear
categories as a category of paths.  Instead of paths, we use the track
objects introduced in the next section. As models, instead
of all d-categories, we take presheaves on a fixed d-category. Using
these notions, we consider open maps in Section~\ref{se:openmaps}
below.

\section{Track objects}
\label{s:tracks}

Track objects allow expressing executions of presheaf automata in
terms of morphisms of presheaf automata. They also allow defining
their languages.  Let again $\cC$ be a d-category. The following
definition is an instance of \eqref{e:Gen}.

\begin{df}
  \label{d:TrackObject}
  The \emph{track object} spanned by a path $\omega:\cI\to\cC$ is the
  simple $\cC$-automaton
  \begin{equation*}
    \widehat{\omega}
    =
    \colim\left(
      \cI\overset\omega\longrightarrow \cC \overset{\yoneda_\cC}\longrightarrow\psh\cC
    \right)
    =
    \colim_{i\in\cI}\; \widehat{\omega(i)}
  \end{equation*}
  with start cell
  $\bot_{\widehat\omega}=\id_{\omega(\bot_\cI)}$ and
  accept cell $\top_{\widehat\omega}=\id_{\omega(\top_\cI)}$.
\end{df}
In terms of cospans \eqref{e:SimpleAutSpan},
$
  \widehat{\omega}
  =
  \big(
  \widehat{\omega(\bot_\cI)}
  \longrightarrow
  \colim_{i\in\cI}\; \widehat{\omega(i)}
  \longleftarrow
  \widehat{\omega(\top_\cI)}
  \big).
$

\def\ms{8pt}
\begin{exa}
  \label{x:GraphTrackObjects}
The representable presheaves in $\cG=\cG(\{a,b\})$ from Section~\ref{s:StdAut} are 
\begin{equation*}
	\widehat{\ast}=
	\begin{tikzpicture}
  \node[state, minimum size=\ms] (0) at (0,0) {};	
  \node[above right] at (0,0) {$\id_\ast$};	
	\end{tikzpicture},
	\qquad
	\widehat{a}=
	\begin{tikzpicture}
  \node[state, minimum size=\ms] (0) at (0,0) {};	
  \node[above left] at (0,0) {$\sigma_a$};	
  \node[state, minimum size=\ms] (1) at (2,0) {};	
  \node[above right] at (2,0) {$\tau_a$};	
	\path (0) edge node {$\id_a$} (1);
	\end{tikzpicture},	
	\qquad
	\widehat{b}=
	\begin{tikzpicture}
  \node[state, minimum size=\ms] (0) at (0,0) {};	
  \node[above left] at (0,0) {$\sigma_b$};	
  \node[state, minimum size=\ms] (1) at (2,0) {};	
  \node[above right] at (2,0) {$\tau_b$};	
	\path (0) edge node {$\id_b$} (1);
	\end{tikzpicture}.
\end{equation*}
The track object $\widehat{\omega}$ of the path 
$\omega:\cI=\upstep*\downstep*\upstep*\downstep*\upstep*\downstep\to\cG$,
\begin{equation*}
  \omega=
  (\ast \arrO{\sigma_{a}} a\arrI{\tau_{a}} 
  \ast \arrO{\sigma_{b}} b\arrI{\tau_{b}}
  \ast \arrO{\sigma_{a}} a\arrI{\tau_{a}}
  \ast),
\end{equation*}
in $\cG$ is thus the colimit of the diagram of $\cG$-presheaves 
  
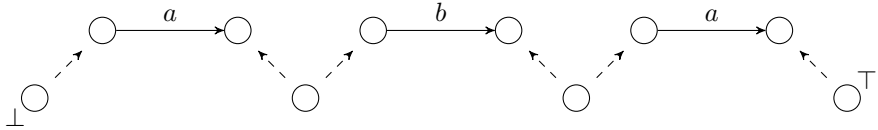
\[
    \begin{tikzpicture}[scale=0.9]
  \node[state, minimum size=\ms] (0) at (0,0) {};
  \node[state, minimum size=\ms] (1) at (4,0) {};
  \node[state, minimum size=\ms] (2) at (8,0) {};
  \node[state, minimum size=\ms] (3) at (12,0) {};
  \node[below left] at (0) {$\bot$};
  \node[above right] at (3) {$\top$};
  \node[state, minimum size=\ms] (a1) at (1,1) {};
  \node[state, minimum size=\ms] (b1) at (3,1) {};
	\path (a1) edge node {$a$} (b1);
  \node[state, minimum size=\ms] (a2) at (5,1) {};
  \node[state, minimum size=\ms] (b2) at (7,1) {};
	\path (a2) edge node {$b$} (b2);
  \node[state, minimum size=\ms] (a3) at (9,1) {};
  \node[state, minimum size=\ms] (b3) at (11,1) {};
	\path (a3) edge node {$a$} (b3);
  	\path (0) edge[dashed, shorten <=0.2cm, shorten >=0.2cm] (a1);
  	\path (1) edge[dashed, shorten <=0.2cm, shorten >=0.2cm] (b1);
  	\path (1) edge[dashed, shorten <=0.2cm, shorten >=0.2cm] (a2);
  	\path (2) edge[dashed, shorten <=0.2cm, shorten >=0.2cm] (b2);
  	\path (2) edge[dashed, shorten <=0.2cm, shorten >=0.2cm] (a3);
  	\path (3) edge[dashed, shorten <=0.2cm, shorten >=0.2cm] (b3);
     \end{tikzpicture}
  \]
 and therefore
  \[
  		\widehat{\omega}=
  \vcenter{\hbox{
  \begin{tikzpicture}
  \node[state, minimum size=\ms] (0) at (0,0) {};
  \node[state, minimum size=\ms] (1) at (2,0) {};
  \node[state, minimum size=\ms] (2) at (4,0) {};
  \node[state, minimum size=\ms] (3) at (6,0) {};
  \path (0) edge node {$a$} (1);
  \path (1) edge node {$b$} (2);
  \path (2) edge node {$a$} (3);
  \node[below left] at (0) {$\bot$};
  \node[above right] at (3) {$\top$};
   \end{tikzpicture}
   }}
\]
In this example, $\cI$ and $\Cell{\widehat\omega}$ are isomorphic
objects in $\Cat\slice\cC$, which illustrates the density theorem
(Lemma \ref{l:Density}).
The track object of
\[
  \eta=(a\arrI{\tau_{a}} \ast \arrO{\sigma_{b}} b\arrI{\tau_{b}} \ast
  \arrO{\sigma_{a}} a)
\]
is isomorphic to $\widehat\omega$ as a $\cG$-presheaf, but has
different start and accept cells (edges instead of vertices):
  \[
  		\widehat{\eta}=
  \vcenter{\hbox{
  \begin{tikzpicture}
  \node[state, minimum size=\ms] (0) at (0,0) {};
  \node[state, minimum size=\ms] (1) at (2,0) {};
  \node[state, minimum size=\ms] (2) at (4,0) {};
  \node[state, minimum size=\ms] (3) at (6,0) {};
  \path (0) edge node {$a$} node[swap] {$\bot$} (1);
  \path (1) edge node {$b$} (2);
  \path (2) edge node {$a$} node[swap] {$\top$} (3);
   \end{tikzpicture}
   }}
  \]
\end{exa}

\begin{exa}
  \label{x:hdapaths2}
  The track objects of the paths $\alpha$, $\beta$, and $\gamma$ of
  Example \eqref{ex:hdapaths} are isomorphic to the $\oldsq$-presheaf
  $X$.  The track object of the path $\zeta$ is isomorphic to
  the sub-presheaf of $X$ consisting of the cells $a$, $p$, $b$, $s$
  and $d$.
\end{exa}

\begin{exa}
\label{ex:TrackObject}
	The track object of the path
	\[
		\omega=
		(
			[0]\arrO{d^0_1}
			[1]\arrO{d^0_2}
			[2]\arrI{d^1_1}
			[1]\arrO{d^0_1}
			[2]\arrI{d^1_{1,2}}
			[0]
		)
	\]
	in $\oldsq$ is the colimit of the diagram
  
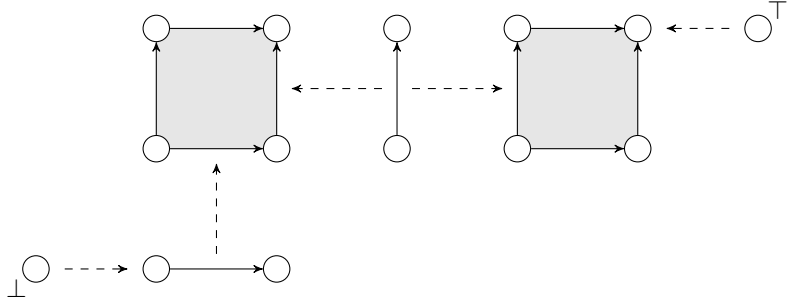
\[
  \begin{tikzpicture}[scale=0.8]
  \node[state, minimum size=\ms] (a) at (0,0) {};
  \node[state, minimum size=\ms] (b0) at (2,0) {};
  \node[state, minimum size=\ms] (b1) at (4,0) {};
  \path (b0) edge (b1);
	\filldraw[color=black!10!white] (2,2)--(4,2)--(4,4)--(2,4)--(2,2);
  \node[state, minimum size=\ms] (c00) at (2,2) {};
  \node[state, minimum size=\ms] (c01) at (2,4) {};
  \node[state, minimum size=\ms] (c10) at (4,2) {};
  \node[state, minimum size=\ms] (c11) at (4,4) {};
  \path (c00) edge (c01);
  \path (c00) edge (c10);
  \path (c01) edge (c11);
  \path (c10) edge (c11);
  \node[state, minimum size=\ms] (d0) at (6,2) {};
  \node[state, minimum size=\ms] (d1) at (6,4) {};
  \path (d0) edge (d1);
	\filldraw[color=black!10!white] (8,2)--(10,2)--(10,4)--(8,4)--(8,2);
  \node[state, minimum size=\ms] (e00) at (8,2) {};
  \node[state, minimum size=\ms] (e01) at (8,4) {};
  \node[state, minimum size=\ms] (e10) at (10,2) {};
  \node[state, minimum size=\ms] (e11) at (10,4) {};
  \path (e00) edge (e01);
  \path (e00) edge (e10);
  \path (e01) edge (e11);
  \path (e10) edge (e11);
  \node[state, minimum size=10pt] (f) at (12,4) {};
  \node[below left] at (a) {$\bot$};
  \node[above right] at (f) {$\top$};
  	\path (a) edge[dashed, shorten <=0.2cm, shorten >=0.2cm] (b0);
  	\path (3,0) edge[dashed, shorten <=0.2cm, shorten >=0.2cm] (3,2);
  	\path (6,3) edge[dashed, shorten <=0.2cm, shorten >=0.2cm] (4,3);
  	\path (6,3) edge[dashed, shorten <=0.2cm, shorten >=0.2cm] (8,3);
  	\path (f) edge[dashed, shorten <=0.2cm, shorten >=0.2cm] (e11);
     \end{tikzpicture}
  \]
  and therefore
\[
	\widehat\omega
	=
  \vcenter
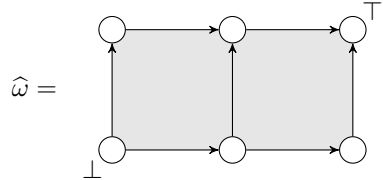
{\hbox{	
  \begin{tikzpicture}[scale=0.8]
	\filldraw[color=black!10!white] (0,0)--(4,0)--(4,2)--(0,2)--(0,0);
  \node[state, minimum size=\ms] (00) at (0,0) {};
  \node[state, minimum size=\ms] (10) at (2,0) {};
  \node[state, minimum size=\ms] (20) at (4,0) {};
  \node[state, minimum size=\ms] (01) at (0,2) {};
  \node[state, minimum size=\ms] (11) at (2,2) {};
  \node[state, minimum size=\ms] (21) at (4,2) {};
  \path (00) edge (10);
  \path (10) edge (20);
  \path (01) edge (11);
  \path (11) edge (21);
  \path (00) edge (01);
  \path (10) edge (11);
  \path (20) edge (21);
  \node[below left] at (00) {$\bot$};
  \node[above right] at (21) {$\top$};
  \end{tikzpicture}
  }}
  \]
\end{exa}

\begin{df}
A \emph{track object} in $\cC$ (a \emph{$\cC$-track object}) is a
simple $\cC$-automaton isomorphic to the track object of some path in
$\cC$.
\end{df}

Let $\tro\cC\subseteq\saut\cC$ denote the full subcategory of track
objects. Morphisms of $\tro\cC$
are called \emph{subsumptions}
\cite{FJSZ_HDALang}. We say
that a track object $\Delta$ \emph{subsumes} $\Gamma$, written
$\Gamma\subsu\Delta$, if there exists a subsumption
$\Gamma\to \Delta$.

\begin{df}
\label{d:TrOConcat}
The \emph{concatenation} of simple $\cC$-automata $X$ and $Y$,
for which $\tgt(X)=\src(Y)=U\in\cC$, is the simple $\cC$-automaton
\begin{equation*}
  X*Y=
  \big(
  \widehat{\src(X)}
  \longrightarrow
  \colim
  \big(
    X
    \longleftarrow
    \widehat{U}
    \longrightarrow
    Y
  \big)
  \longleftarrow
  \widehat{\tgt(Y)}
  \big)
  .
\end{equation*}
\end{df}

\begin{lem}
\label{l:SAutAssociativity}
Concatenation of simple automata is associative, 
\begin{equation*}
  (X*Y)*Z\cong X*(Y*Z),
\end{equation*}
for all $X,Y,Z\in\saut\cC$ such that $\tgt(X)=\src(Y)$ and
$\tgt(Y)=\src(X)$.
\end{lem}
\begin{proof}
  This follows from the universal property of pushouts \cite[Example
  2.1.22]{Johnson-Yau_2Categories}.
\end{proof}

\begin{lem}
\label{l:PathComposition}
For paths $\omega, \eta$ in $\cC$
with $\tgt(\omega)=\src(\eta)$,
there is a natural isomorphism
\[
	\widehat{\omega*\eta}
	\cong
	\widehat\omega*\widehat\eta.
\]
\end{lem}
\begin{proof}
	This follows from the universal property of colimits.
\end{proof}
The following consequence is immediate.
\begin{cor}
  Concatenations of track objects are track objects.
\end{cor}

Such concatenations are therefore associative up to natural
isomorphism.

\begin{exa}
  For
  $\omega=		(
  [0]\arrO{d^0_{1,2}}
  [2]\arrI{d^1_1}
  [1]
  )
  $ and
  $\eta=		(
  [1]\arrI{d^1_1}
  [0]\arrO{d^1_1}
  [1]
  )
  $,
  \[
    \begin{tikzpicture}[scale=0.9]
      \filldraw[color=black!10!white] (0,0)--(2,0)--(2,2)--(0,2)--(0,0);
      \node at (1,3) {$\widehat{\omega}$};
      \node[state, minimum size=\ms] (00) at (0,0) {};
      \node[state, minimum size=\ms] (10) at (2,0) {};
      \node[state, minimum size=\ms] (01) at (0,2) {};
      \node[state, minimum size=\ms] (11) at (2,2) {};
      \path (00) edge (10);
      \path (01) edge (11);
      \path (00) edge (01);
      \path (10) edge (11);
      \node[below left] at (00) {$\bot$};
      \node[right] at (2,1) {$\top$};
      \node at (3,3) {$*$};
      \node at (3,1) {$*$};
      \node at (5,3) {$\widehat{\eta}$};
      \node[state, minimum size=\ms] (a00) at (4,0) {};
      \node[state, minimum size=\ms] (a01) at (4,2) {};
      \node[state, minimum size=\ms] (a11) at (6,2) {};
      \path (a01) edge (a11);
      \path (a00) edge (a01);
      \node[left] at (4,1) {$\bot$};
      \node[above] at (5,2) {$\top$};
      \node at (7,3) {$=$};
      \node at (7,1) {$=$};
      \filldraw[color=black!10!white] (8,0)--(8,2)--(10,2)--(10,0)--(8,0);
      \node at (9,3) {$\widehat{\omega}*\widehat{\eta}=\widehat{\omega*\eta}$};
      \node[state, minimum size=\ms] (b00) at (8,0) {};
      \node[state, minimum size=\ms] (b10) at (10,0) {};
      \node[state, minimum size=\ms] (b01) at (8,2) {};
      \node[state, minimum size=\ms] (b11) at (10,2) {};
      \node[state, minimum size=\ms] (b21) at (12,2) {};
      \path (b00) edge (b10);
      \path (b01) edge (b11);
      \path (b00) edge (b01);
      \path (b10) edge (b11);
      \path (b11) edge (b21);
      \node[below left] at (b00) {$\bot$};
      \node[above] at (11,2) {$\top$};
    \end{tikzpicture}
  \]
\end{exa}

\begin{df}
  \label{d:ConcatSubsu}
  Let $U,V,W\in\Ob(\cC)$, $\Gamma,\Gamma'\in\ilo{U}{\tro\cC}{V}$ and
  $\Delta,\Delta'\in\ilo{V}{\tro\cC}{W}$.  The \emph{concatenation} of
  subsumptions $f:\Gamma\to\Gamma'$ and $g:\Delta\to\Delta'$ is the
  subsumption $f*g:\Gamma*\Delta\to \Gamma'*\Delta'$ defined by the
  diagram
  \[
    \begin{tikzcd}[row sep=tiny]
      & &
      \Gamma*\Delta
      \arrow[bend left=40]{dddd}[right]{f*g}
      & &
      \\
      &
      \Gamma
      \arrow{ur}
      \arrow{dd}[left]{f}
      & &
      \Delta
      \arrow{ul}
      \arrow{dd}[right]{g}
      &
      \\
      \widehat{U}
      \arrow{ur}
      \arrow{dr}&  &
      \widehat{V}
      \arrow{ul}
      \arrow{dl}
      \arrow{ur}
      \arrow{dr}
      & &
      \widehat{W}\arrow{ul} \arrow{dl}
      \\
      &
      \Gamma'
      \arrow{dr}
      & &
      \Delta'
      \arrow{dl}
      &
      \\
      & & \Gamma'*\Delta' & 	&
    \end{tikzcd}
  \]
\end{df}

\begin{cor}
\label{c:ConcatSubsu}
The subsumption relation is a precongruence: if
$\Gamma\subsu\Gamma'$ and $\Delta\subsu\Delta'$, then
$\Gamma*\Delta\subsu\Gamma'*\Delta'$.  
\end{cor}

A track object is \emph{elementary} if it is isomorphic to a track
object of a step.  Elementary track objects thus have the form
\begin{equation*}
( \widehat{U}\overset{\id}\longrightarrow\widehat{U}
  \overset{\widehat{\psi}}\longleftarrow \widehat{V})
\end{equation*}
for $\phi\in\forM\cC(V,U)$ or $\psi\in\bckM\cC(V,U)$.
By Lemma \ref{l:PathComposition}, every track
object is isomorphic to a concatenation of elementary track objects.
The concatenation may be empty: the track object associated with the
constant path $(U)$, $U\in\Ob(\cC)$, is the \emph{identity track object}
$\idtrack{U}=(\widehat{U}\xrightarrow\id \widehat{U}\xleftarrow\id \widehat{U})$.

\begin{lem}
\label{l:TrackObjectGens}
A simple $\cC$-automaton is a track object if and only if it is
isomorphic to a concatenation of elementary track objects.
\end{lem}
\begin{proof}
	This follows from Lemmas \ref{l:PathConcatSteps} and \ref{l:PathComposition}.
\end{proof}

\begin{prp}
  \label{p:TrOPathEq}
  	Track objects of equivalent paths in $\cC$ are isomorphic.  
\end{prp}

\begin{proof}
  We need to check that the relations from Lemma \ref{l:ClRefin} preserve track objects.
  For $\varphi\in\forM\cC(U,V)$, $\psi\in\forM\cC(V,W)$,
  \begin{align*}
    \widehat{(W\arrO\psi V)}*\widehat{(V\arrO\varphi U)}
    &=
    (\widehat{W}\xrightarrow{\psi}\widehat{V}\xleftarrow\id \widehat{V})
    *
    (\widehat{V}\xrightarrow{\varphi}\widehat{U}\xleftarrow\id \widehat{U})		
    \\
    &\cong
    (\widehat{W}\xrightarrow{\psi\circ \varphi}\widehat{U}\xleftarrow\id \widehat{U})	\\	
    &\cong
    \widehat{(W\arrO{\varphi\circ \psi}U)},
  \end{align*}
  shows relation (1) in Lemma \ref{l:ClRefin}. Track objects are
  presented here as cospans as in \eqref{e:SimpleAutSpan}.
  Verification of the remaining relations is similar.
\end{proof}

\def\tg{\mathcal{T}(\Sigma)}

In the next lemma, we describe track objects in the category $\cG(\Sigma)$.
Let $\tg$ be the discrete category given by the following data:
\begin{itemize}
    \item objects are strings
$b_uwe_u$, where $w\in\Sigma^*$, $b_u\in\{\varepsilon\}\cup\{\tau_a\}_{a\in\Sigma}$ and $e_u\in\{\varepsilon\}\cup\{\sigma_a\}_{a\in\Sigma}$,
\item 
    the source of $b_uwe_v$ is $\vo$ if $b_u=\varepsilon$
    and $a$ if $b_u=\tau_a$,
\item 
    the target of $b_uwe_v$ is $\vo$ if $e_v=\varepsilon$
    and $a$ if $e_v=\tau_a$,
\item 
    concatenation of objects with matching sources and targets is string concatenation with the convention $\sigma_a\tau_a=a$ for $a\in\Sigma$.
\end{itemize}
\begin{lem}
\label{l:TracksOfG}
    The category of track objects $\tro{\cG}$ on $\cG=\cG(\Sigma)$
    is equivalent to $\tg$.
    In particular, the set $\ilo{\vo}{\tro{\cG}}{\vo}$ of isomorphism classes of $\cG$-track objects $\Gamma$ satisfying $\src(\Gamma)=\tgt(\Gamma)=\vo$ is isomorphic to the free monoid $\Sigma^*$.
\end{lem}
\begin{proof}
    Every path in $\cG$ is isomorphic to an alternating sequence of steps $\sigma_a$ and $\tau_b$ ($a,b\in\Sigma$), similarly to Example \ref{x:PathsInG}, where every $\sigma_a$ is followed by $\tau_a$. We merge $\sigma_a\tau_a$ into the letter $a$.
    As presented in Example \ref{x:GraphTrackObjects},
    track objects are ``linear'' automata;  every morphism between them is an isomorphism. Concatenation is clear in light of Lemma \ref{l:PathComposition}.
\end{proof}

We are now ready to show that track objects provide
an alternative categorical notion of execution of presheaf
automata.

\begin{df}
  \label{d:Track}
  A \emph{track} in a $\cC$-presheaf $X$ is a presheaf map
  $\boldsymbol\alpha: \Gamma\to X$ from a track object~$\Gamma$.  The
  \emph{source} of a track $\boldsymbol\alpha:\Gamma\to X$ is the cell
  $\src(\boldsymbol\alpha)=\boldsymbol\alpha(\bot_\Gamma)$, its
  \emph{target} is
  $\tgt(\boldsymbol\alpha)=\boldsymbol\alpha(\top_\Gamma)$.  A track
  $\boldsymbol\alpha:\Gamma\to X$ in a $\cC$-automaton $X$ is
  \emph{accepting} if $\boldsymbol\alpha$ is a morphism of
  $\cC$-automata, that is, $\boldsymbol\alpha(\bot_\Gamma)\in \bot_X$
  and $\boldsymbol\alpha(\top_\Gamma)\in \top_X$.
\end{df}

\begin{df}
The \emph{concatenation} of tracks $\boldsymbol\alpha:\Gamma\to X$
and $\boldsymbol\beta:\Delta\to X$
such that $\tgt(\boldsymbol\alpha)=\src(\boldsymbol\beta)\in X[U]$
is the track
\[
  \boldsymbol\alpha*\boldsymbol\beta:
  \Gamma*\Delta
  =
  \colim(\Gamma \xleftarrow{\top_\Gamma} \widehat{U} \xrightarrow{\bot_\Delta} \Delta)
  \xrightarrow{\boldsymbol\alpha\cup \boldsymbol\beta}
  X.
\]
\end{df}

Track concatenation is associative up to natural isomorphism:
\[
  (\boldsymbol\alpha*\boldsymbol\beta)*\boldsymbol\gamma\cong
  \boldsymbol\alpha*(\boldsymbol\beta*\boldsymbol\gamma).
  \]

\begin{rem}
  The above results can be extended to show that track objects in
  $\cC$ (or simple $\cC$-automata more generally) form a bicategory
  with objects of $\cC$ as $0$-cells, track objects as $1$-cells and
  subsumptions as 2-cells.  Composition of $1$-cells as well as
  horizontal composition of $2$-cells (Definition \ref{d:ConcatSubsu})
  is concatenation (Definition \ref{d:TrOConcat}), while vertical
  composition of $2$-cells is the composition of subsumptions. This
  forms a sub-bicategory of the cospan bicategory of $\psh\cC$.
\end{rem}

Functors between d-categories translate track objects on the source category into those on the target category.
Fix a d-functor $F:\cD\to\cC$.
\begin{df}\ The \emph{direct image} of a simple $\cD$-automaton
  \hbox{$\Delta=(\widehat{V}\to X\gets \widehat{W})$} along $F$ is the simple
  $\cC$-automaton
  \[
    F_*\Delta=(\widehat{F(V)}=F_*\widehat{V}\to F_*X \gets F_*\widehat{W}=\widehat{F(W)}).
  \]
\end{df}

\begin{lem}
  \label{l:DirectImageOfPath}
  Let $\eta:\cI\to \cD$ be a path. Then
  $F_*\widehat{\eta}\cong \widehat{F\eta}$.
\end{lem}
\begin{proof}
  $
    \widehat{F\eta}
    =
    \colim_{i\in \cI} \widehat{(F\eta)(i)}
    =
    \colim_{i\in \cI} F_*\widehat{\eta(i)}
    =
    F_*\colim_{i\in \cI} \widehat{\eta(i)}
    =
    F_*\widehat{\eta}.
  $
\end{proof}

\begin{cor}
    The direct image of a track object is a track object.
\end{cor}
\begin{lem}
\label{l:BasicPropertiesOfDirectImages}
Suppose $\Gamma$ and $\Delta$ are track objects in $\cD$. If $\tgt(\Gamma)=\src(\Delta)$, then $F_*(\Gamma*\Delta)\cong F_*\Gamma*F_*\Delta$, and if $\Delta\subsu\Gamma$, then $F_*\Delta\subsu F_*\Gamma$.
\end{lem}
\begin{proof}
    Since direct images preserve colimits, the first statement is a consequence of Definition \ref{d:TrOConcat}. The second follows from the functoriality of $F_*$.
\end{proof}

\section{Languages of presheaf automata}

In this section, we define languages of presheaf automata using track objects and show how accepting paths and tracks are related. To avoid set-theoretic complications, we pass to isomorphism classes of tracks.

\def\Words{\mathscr{W}}
\begin{df}
    A \emph{$\cC$-word} is an isomorphism class of $\cC$-track objects. The set of $\cC$-words is denoted $\Words(\cC)$. 
\end{df}

As isomorphism classes of linear categories form a set and $\cC$ is small, $\Words(\cC)$ is indeed a set.

\begin{df}
  The \emph{language} of a $\cC$-automaton $X$ is the set of all equivalence classes of track objects of accepting tracks:
  \[
    \Lang(X)=\{[\Gamma]\in \Words(\cC)\mid \aut\cC(\Gamma,X)\neq \emptyset\}.
  \]
\end{df}

\begin{rem}
    If convenient, we regard languages of automata as families of track objects rather than their isomorphism classes and write $\Gamma\in\Lang(X)$ instead of $[\Gamma]\in\Lang(X)$.
\end{rem}

\begin{lem}
\label{l:LanguageOfTrackObject}
    If $\Gamma\in\tro\cC$, then $\Lang(\Gamma)=\{[\Delta]\in\Words(\cC)\mid \Delta\subsu\Gamma\}$.
\end{lem}
\begin{proof}
    This is follows immediately from the definition of subsumption.
\end{proof}

\begin{lem}
\label{l:LangFun}
\label{l:ImageOfLanguage}
	If $f:Y\to X$ is a map of\/ $\cC$-automata, then $\Lang(Y)\subseteq \Lang(X)$.
\end{lem}

\begin{proof}
	If $[\Gamma]\in \Lang(Y)$, then there exists a map $\boldsymbol\alpha:\Gamma\to Y$.
	Thus, $f \circ \boldsymbol\alpha$ is a map $\Gamma\to X$ and hence $[\Gamma]\in \Lang(X)$.
\end{proof}

\begin{lem}
\label{l:LanguageOfCoproduct}
$\Lang(\coprod_{i\in I}X_i)=\bigcup_{i\in I} \Lang(X_i)$
for every family $\{X_i\}_{i\in I}$ of $\cC$-automata.
\end{lem}
\begin{proof}
  For every $j\in I$, 
  $\Lang(X_j)\subseteq \Lang(\coprod_{i\in I}X_i)$ follows from Lemma \ref{l:LangFun} and therefore 
    $\bigcup_{i\in I} \Lang(X_i) \subseteq \Lang(\coprod_{i\in I}X_i)$.
  Every track object $\Gamma$ is connected as a colimit of   connected presheaves on a connected category. Hence, every   morphism in $\aut{\cC}(\Gamma,\coprod_{i\in I}X_i)$ factors through $X_j$ for some $j\in I$ and $\Lang(\coprod_{i\in I}X_i)\subseteq\bigcup_{i\in I} \Lang(X_i)$.
\end{proof}

\begin{exa}
  In the formalism of presheaf automata, the language of a standard automaton is formed by ``linear'' automata rather than words, see Lemma \ref{l:TracksOfG} for the construction. The language of the standard automaton
  from Example \ref{e:StdAutExample}, for instance, is
  \begin{equation*}
    \left\{{\cdot}{\tto{a}}{\cdot},
    {\cdot}{\tto{a}}{\cdot}{\tto{b}}{\cdot},
    {\cdot}{\tto{a}}{\cdot}{\tto{b}}{\cdot}{\tto{a}}{\cdot},
    {\cdot}{\tto{a}}{\cdot}{\tto{b}}{\cdot}{\tto{b}}{\cdot},
    {\cdot}{\tto{a}}{\cdot}{\tto{b}}{\cdot}{\tto{a}}{\cdot}{\tto{b}}{\cdot},
    {\cdot}{\tto{a}}{\cdot}{\tto{b}}{\cdot}{\tto{b}}{\cdot}{\tto{b}}{\cdot},
    \dotsc\right\}.
  \end{equation*}
\end{exa}

\begin{exa}
\label{x:PrecubeLanguage}
Languages of $\oldsq$-automata can be seen as sets of ``shapes''.  The
language of the $\oldsq$-automaton from Example \ref{x:oldsq1}, for
instance, is
\begin{equation*}
  \left\{
  \vcenter{\hbox{%
      \begin{tikzpicture}[scale=.7, -, transform shape]
        \filldraw[color=black!10!white] (0,0)--(1,0)--(1,1)--(0,1)--(0,0);
        \node[state] (00) at (0,0) {};
        \node[state] (10) at (1,0) {};
        \node[state] (01) at (0,1) {};
        \node[state] (11) at (1,1) {};
        \path (00) edge (10);
        \path (01) edge node[above] {$\top$} (11);
        \path (00) edge (01);
        \path (10) edge (11);
        \node[below] at (00) {$\bot$};
      \end{tikzpicture}}},
  \vcenter{\hbox{%
      \begin{tikzpicture}[scale=.7, -, transform shape]
        \node[state] (00) at (0,0) {};
        \node[state] (01) at (0,1) {};
        \node[state] (11) at (1,1) {};
        \path (01) edge node[above] {$\top$} (11);
        \path (00) edge (01);
        \node[below] at (00) {$\bot$};
      \end{tikzpicture}}},
  \vcenter{\hbox{%
      \begin{tikzpicture}[scale=.7, -, transform shape]
        \filldraw[color=black!10!white] (0,0)--(1,0)--(1,1)--(0,1)--(0,0);
        \node[state] (00) at (0,0) {};
        \node[state] (10) at (1,0) {};
        \node[state] (01) at (0,1) {};
        \node[state] (11) at (1,1) {};
        \node[state] (21) at (2,1) {};
        \path (00) edge (10);
        \path (01) edge (11);
        \path (11) edge (21);
        \path (00) edge (01);
        \path (10) edge (11);
        \node[below] at (00) {$\bot$};
        \node[above] at (21) {$\top$};
      \end{tikzpicture}}},
  \vcenter{\hbox{%
      \begin{tikzpicture}[scale=.7, -, transform shape]
        \node[state] (00) at (0,0) {};
        \node[state] (10) at (1,0) {};
        \node[state] (11) at (1,1) {};
        \node[state] (21) at (2,1) {};
        \path (00) edge (10);
        \path (11) edge (21);
        \path (10) edge (11);
        \node[below] at (00) {$\bot$};
        \node[above] at (21) {$\top$};
      \end{tikzpicture}}}
  \right\}.
\end{equation*}
Note that there are two different tracks from the right-most track object.

Track objects in $\oldsq$ correspond to \emph{pomsets with interfaces} (\emph{ipomsets}) over a
one-letter alphabet \cite{FJSZ_HDALang}.  An ipomset is a
tuple $(P,{<},{\intord},S,T)$, where $P$ is a finite set, $<$ and
$\intord$ are partial orders on $P$ and $S,T\subseteq P$ are subsets, called \emph{interfaces}, 
of the sets of $<$-minimal and $<$-maximal elements of $P$,
respectively (compared to \cite{FJSZ_HDALang} we disregard the
labelling function).  In terms of ipomsets, the above language is
\begin{equation*}
    \left\{\;
      \vcenter{\hbox{%
          \begin{tikzpicture}[scale=.7, -, transform shape]
            \node[state] (A) at (0,1) {};
            \node[state] (B) at (0,0) {};
            \path[->,very thin,dashed] (A) edge (B);
            \node[right] at (A) {$\top$};
          \end{tikzpicture}}},\quad
      \vcenter{\hbox{%
          \begin{tikzpicture}[scale=.7, -, transform shape]
            \node[state] (A) at (0,0.5) {};
            \node[state] (B) at (1,0.5) {};
            \path[->] (A) edge (B);
            \node[right] at (B) {$\top$};
          \end{tikzpicture}}},\quad
      \vcenter{\hbox{%
          \begin{tikzpicture}[scale=.7, -, transform shape]
            \node[state] (A) at (0,1) {};
            \node[state] (B) at (0,0) {};
            \node[state] (C) at (1,0.5) {};
            \path[->] (A) edge (C);
            \path[->] (B) edge (C);
            \path[->,very thin, dashed] (A) edge (B);
          \end{tikzpicture}}},\quad
      \vcenter{\hbox{%
          \begin{tikzpicture}[scale=.7, -, transform shape]
            \node[state] (A) at (0,0.5) {};
            \node[state] (B) at (1,0.5) {};
            \node[state] (C) at (2,0.5) {};
            \path[->] (A) edge (B);
            \path[->] (B) edge (C);
          \end{tikzpicture}}}
    \right\}.
\end{equation*}
Elements of $T$ are marked by $\top$, and all the sets $S$ are empty.
\end{exa}

Alternatively, languages can be described using accepting paths. \begin{df}
  The \emph{canonical section} of the path $\omega:\cI\to\cC$ is the path $\kappa_\omega:\cI\to\Cell{\widehat\omega}$ given by the unit of the adjunction between the categories
$\psh\cC$ and $\dCat\slice\cC$ from Proposition \ref{p:dAdj}.
\end{df}

\begin{exa}
  Consider again the path
  \begin{equation*}
    \omega= ( [0]\arrO{d^0_1} [1]\arrO{d^0_2} [2]\arrI{d^1_1}
    [1]\arrO{d^0_1} [2]\arrI{d^1_{12}} [0] )
  \end{equation*}
  from Example \ref{ex:TrackObject}.  The following picture shows the
  canonical section $\kappa_\omega$ on $\widehat\omega$ on the left
  and values of $\widehat{\omega}$ on the right:
	
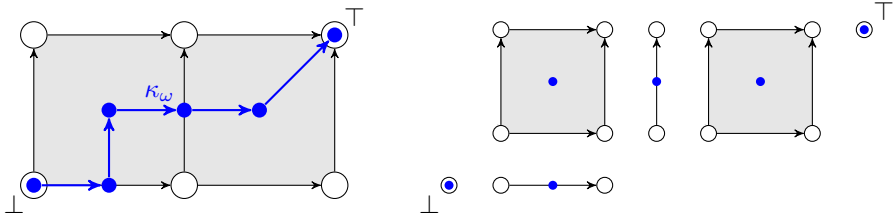
\[
  \begin{tikzpicture}
	\filldraw[color=black!10!white] (0,0)--(4,0)--(4,2)--(0,2)--(0,0);
  \node[state, minimum size=10pt] (00) at (0,0) {};
  \node[state, minimum size=10pt] (10) at (2,0) {};
  \node[state, minimum size=10pt] (20) at (4,0) {};
  \node[state, minimum size=10pt] (01) at (0,2) {};
  \node[state, minimum size=10pt] (11) at (2,2) {};
  \node[state, minimum size=10pt] (21) at (4,2) {};
  \path (00) edge (10);
  \path (10) edge (20);
  \path (01) edge (11);
  \path (11) edge (21);
  \path (00) edge (01);
  \path (10) edge (11);
  \path (20) edge (21);
  \node[above left] at (2,1) {\color{blue}$\kappa_\omega$};
  \node[below left] at (00) {$\bot$};
  \node[above right] at (21) {$\top$};
   \node[circle, inner sep=2pt, fill=blue] (a00) at (0,0) {};
  \node[circle, inner sep=2pt, fill=blue] (a10) at (1,0) {};
  \node[circle, inner sep=2pt, fill=blue] (a11) at (1,1) {};
  \node[circle, inner sep=2pt, fill=blue] (a21) at (2,1) {};
  \node[circle, inner sep=2pt, fill=blue] (a31) at (3,1) {};
  \node[circle, inner sep=2pt, fill=blue] (a42) at (4,2) {};
  \path (a00) edge[blue,thick] (a10);
  \path (a10) edge[blue,thick] (a11);
  \path (a11) edge[blue,thick] (a21);
  \path (a21) edge[blue,thick] (a31);
  \path (a31) edge[blue,thick] (a42);
  \begin{scope}[scale=0.69, shift={(8,0)}]
  \node[state, minimum size=6pt] (a) at (0,0) {};
  \node[state, minimum size=6pt] (b0) at (1,0) {};
  \node[state, minimum size=6pt] (b1) at (3,0) {};
  \path (b0) edge (b1);
	\filldraw[color=black!10!white] (1,1)--(3,1)--(3,3)--(1,3)--(1,1);
  \node[state, minimum size=6pt] (c00) at (1,1) {};
  \node[state, minimum size=6pt] (c01) at (1,3) {};
  \node[state, minimum size=6pt] (c10) at (3,1) {};
  \node[state, minimum size=6pt] (c11) at (3,3) {};
  \path (c00) edge (c01);
  \path (c00) edge (c10);
  \path (c01) edge (c11);
  \path (c10) edge (c11);
  \node[state, minimum size=6pt] (d0) at (4,1) {};
  \node[state, minimum size=6pt] (d1) at (4,3) {};
  \path (d0) edge (d1);
	\filldraw[color=black!10!white] (5,1)--(7,1)--(7,3)--(5,3)--(5,1);
  \node[state, minimum size=6pt] (e00) at (5,1) {};
  \node[state, minimum size=6pt] (e01) at (5,3) {};
  \node[state, minimum size=6pt] (e10) at (7,1) {};
  \node[state, minimum size=6pt] (e11) at (7,3) {};
  \path (e00) edge (e01);
  \path (e00) edge (e10);
  \path (e01) edge (e11);
  \path (e10) edge (e11);
  \node[state, minimum size=6pt] (f) at (8,3) {};
  \node[below left] at (a) {$\bot$};
  \node[above right] at (f) {$\top$};
   \node[circle, inner sep=1.2pt, fill=blue] at (0,0) {};
   \node[circle, inner sep=1.2pt, fill=blue] at (2,0) {};
   \node[circle, inner sep=1.2pt, fill=blue] at (2,2) {};
   \node[circle, inner sep=1.2pt, fill=blue] at (4,2) {};
   \node[circle, inner sep=1.2pt, fill=blue] at (6,2) {};
   \node[circle, inner sep=1.2pt, fill=blue] at (8,3) {};
   \end{scope}
 \end{tikzpicture}	
\]
\end{exa}

For a $\cC$-presheaf $X$ let
$\Path(X):=\Path(\Cell{X})$ denote the category of paths in $X$.  For a path $\omega:\cI\to \cC$, the set of all paths with ``shape'' $\omega$ on $X$ is defined as
\begin{align}
  \Path(X;\omega) \label{e:PathShape}
  &=\{\alpha\in\dCat(\cI,\Cell X)\mid \pi_x\circ \alpha=\omega\}\\
  &\cong \{\alpha\in\Cat(\cI,\Cell X)\mid \pi_x\circ \alpha=\omega\}\notag \\
  &\cong (\Cat{\downarrow}\cC)((\cI,\omega), (\Cell X,\pi_X)).\notag
\end{align}
The middle equation holds because the d-structure on $\Cell X$ is induced by the projection on $\cC$.

\begin{lem}
  \label{l:PathTrackCorrespondence}
  For a path $\omega\in\dCat(\cI,\cC)$ and $X\in\psh\cC$ the maps
 \begin{align*}
 	\Path(X;\omega)
 	\ni
 	\alpha
 	&\mapsto
 	(\widehat{\pi_X\circ \alpha}
	\xrightarrow{\alpha_*}
	\widehat\pi_X
	\xrightarrow\cong
	X)
	\in\psh\cC(\widehat\omega,X),
 \\
    \psh\cC(\widehat\omega,X)
    \ni
    \boldsymbol\alpha
    &\mapsto
    \Cell(\boldsymbol\alpha)\circ \kappa_\omega
    \in
    \Path(X;\omega)
  \end{align*}
  are inverses.
\end{lem}

\begin{proof}
This is an immediate consequence of the adjunction \eqref{e:CellGenAdj}.
\end{proof}

We can now show that languages of $\cC$-automata can be computed from
their accepting paths.

\begin{cor}
  For every $X\in\aut\cC$,
  \[
    \Lang(X)=\{[\widehat{\pi_X\circ\alpha}]\mid\text{$\alpha$ is an accepting path in $X$} \}.
  \]
\end{cor}

\begin{rem}
  Lemma \ref{l:PathTrackCorrespondence} exhibits a bijection between
  $\omega$-shaped paths on $X$ and tracks $\widehat\omega\to X$.  Yet
  there is no equivalence between all paths and all tracks on $X$: a
  track object $\Gamma\in\TrO(\cC)$ can be generated by different
  paths in $\cC$ ($\Gamma\simeq\widehat\omega\simeq\widehat\eta$ for
  $\omega\not\simeq\eta$), in which case a track
  $\boldsymbol\alpha:\Gamma\to X$ corresponds to different paths, 
  namely
  $\Cell(\boldsymbol\alpha)\circ\kappa_\omega\neq
  \Cell(\boldsymbol\alpha)\circ\kappa_\eta$.  Example
  \ref{x:hdapaths2}, for instance, provides three different paths $\alpha$, $\beta$ and $\gamma$ with isomorphic tracks.
\end{rem}

\section{Languages over d-categories}
\label{se:lang}

In this section, we introduce the notion of language parametrised by a d-category $\cC$.  For $\cC=\cG(\Sigma)$ from Section \ref{s:StdAut},
we recover the classical notion.

First, we add structure to the set $\Words(\cC)$ of $\cC$-words. The source, target, concatenation and subsumption of $\cC$-words are defined as for track objects: $\src([\Gamma])=\src(\Gamma)$, $\tgt([\Gamma])=\tgt(\Gamma)$, $[\Gamma]*[\Delta]=[\Gamma*\Delta]$ and $[\Gamma]\subsu[\Delta]$ if and only if $\Gamma\subsu\Delta$. These definitions do not depend on the choice of representatives.

These operations turn $\Words(\cC)$ into a category enriched in the category of posets:
\begin{itemize}
    \item $\Ob(\Words(\cC))=\Ob(\cC)$,
    \item $\Words(\cC)(V,U)=\{w\in\Words(\cC)\mid \src(w)=V,\; \tgt(w)=U\}$ for $U,V\in \Ob(\cC)$,
    \item $\id_U=[\idtrack{U}]$ for $U\in\Ob(\cC)$,
    \item $w\circ v=v*w$ for $v,w\in \Words(\cC)$, $\tgt(v)=\src(w)$,
    \item the partial order on homsets is given by subsumption $\subsu$.
\end{itemize}
Composition is associative by Lemma \ref{l:SAutAssociativity}; it preserves subsumption by Corollary \ref{c:ConcatSubsu}.
Whenever we regard $\Words(\cC)$ as a set, we regard it as the set of morphisms, not the set of objects.

For a subset $A\subseteq \Ob(\cC)$, $\Words_A(\cC)$ 
denotes the full subcategory of $\cC$ with objects $A$.

\begin{exa}
    Lemma \ref{l:TracksOfG} shows that $\Words_{*}(\cG(\Sigma))$ is isomorphic to the free monoid $\Sigma^*$, regarded as a category with a single object and the trivial subsumption order.
\end{exa}

Every d-functor $F:\cD\to\cC$ induces a functor 
\begin{equation}
\label{e:WordsFunctor}
    \Words(F):\Words(\cD)\ni [\Gamma]\mapsto [F_*\Gamma] \in \Words(\cC).
\end{equation}
By Lemma \ref{l:BasicPropertiesOfDirectImages}, it preserves the subsumption order.

\begin{df}
A \emph{$\cC$-language} is a subset $L\subseteq\Words(\cC)$
that is down-closed with respect to subsumption: $v\in L$ and $w\subsu v$ imply $w\in L$ for all $v,w\in \Words(\cC)$. We write $\Langs(\cC)$ for the set of all $\cC$-languages. 
\end{df}

As for languages of $\cC$-automata, we occasionally regard languages as classes of track objects instead of sets of $\cC$-words.

\begin{df}
The \emph{down-closure} of $A\subseteq \Words(\cC)$ is the $\cC$-language
\[
  A{\downarrow} = \{v\in \Words(\cC)\mid v\subsu w \text{ for some } w \in A\}.
\]
\end{df}

The following lemmas show that $\cC$-automata recognise exactly
the $\cC$-languages.

\begin{lem}
	If $X$ is a $\cC$-automaton, then $\Lang(X)$ is a $\cC$-language.
\end{lem}

\begin{proof}
  We need to check that $\Lang(X)$ is down-closed.  If
  $[\Gamma]\in\Lang(X)$ and $\Delta\subsu \Gamma$, then there are maps of $\cC$-automata $f:\Gamma\to X$ and $g:\Delta\to\Gamma$.  So $g\circ f\in\aut\cC(\Delta,X)\neq\emptyset$
  and thus $[\Delta]\in\Lang(X)$.
\end{proof}

%

\begin{lem}
  Every $\cC$-language $L$ is accepted by some $\cC$-automaton.
\end{lem}
\begin{proof}
  For every $w\in L$ choose its representative $\Gamma_w$ and let $X=\coprod_{w\in L}\Gamma_w$, with the coproduct taken in $\aut\cC$ (not in $\saut\cC$).  Then, by Lemmas \ref{l:LanguageOfTrackObject} and \ref{l:LanguageOfCoproduct},
  \[
  	\Lang(X)
  	=
  	\bigcup_{w\in L}\Lang(\Gamma_w)
  	=
  	\bigcup_{w\in L}\{[\Gamma_w]\}{\downarrow}
  	=
  	\bigcup_{w\in L}\{w\}{\downarrow}
  	=
        L{\downarrow}
        =
  	L.
  	\qedhere
  \]
\end{proof}

\begin{lem}
\label{l:CompleteLattice}
   $\Langs(\cC)$ forms a complete distributive lattice with respect to set union.
\end{lem}
\begin{proof}
  Arbitrary unions (and intersections) of $\cC$-languages are again
  $\cC$-languages.  Thus, $\Langs(\cC)$ is a complete sublattice of
  the powerset lattice of $\Words(\cC)$.
\end{proof}

\begin{rem}
  $\cC$-Languages are not necessarily closed under complementation
  because of down-closure. An
  exception are presheaf automata with trivial subsumption on track objects, such as $\cG$-automata.
\end{rem}

Let $\idlang{\cC}=\{[\idtrack{U}]\mid U\in\Ob(\cC)\}\dcl$.
We introduce the following rational operations on~$\Langs(\cC)$:
\begin{itemize}
\item  the union $L+M=L\cup M$,
\item   the concatenation $L* M =
  \{v*w\mid v\in L,\; w\in M,\; \tgt(v)=\src(w)\}\dcl$,
\item the Kleene plus $L^+ = \bigcup_{n\geq 1} L^i$, for $L^1=L$
  and $L^{k+1}=L*L^k$.
\end{itemize}
We introduce the Kleene plus instead of the more standard Kleene star for reasons explained below.
The down-closure of concatenations is necessary: see
\cite{FahrenbergJSZ24} for an example.

\begin{lem}
  \label{l:KleeneUnit}
  $L*\idlang{\cC}=L=\idlang{\cC}*L$ for every $L\in\Langs(\cC)$.
\end{lem}

\begin{proof}
  If $[\Gamma]\in L$, then 
  $[\Gamma]=[\Gamma*\idtrack{\tgt(\Gamma)}]\in L*\idlang{\cC}$.  If $[\Gamma]\in L*\idlang{\cC}$, then there are $[\Delta]\in L$ and
  $\Theta\subsu\idtrack{\tgt(\Gamma)}$ such that
  $\Gamma\subsu \Delta*\Theta$. Thus, by Corollary
  \ref{c:ConcatSubsu},
  \[
    \Gamma
    \subsu		
    \Delta*\Theta
    \subsu		
    \Delta*\idtrack{\tgt(\Gamma)}
    =
    \Delta*\idtrack{\tgt(\Delta)}
    =
    \Delta
  \]
  and then $[\Gamma]\in L$.
\end{proof}

For the next proposition, recall that a quantale $(Q,\le,\cdot)$ is
a complete lattice $(Q,\le)$ equipped with a semigroup structure
$(Q,\cdot)$ whose multiplication preserves all suprema in both
arguments. A quantale homomorphism preserves multiplication and all sups. A quantale is distributive if the underlying lattice is. A
Kleene plus can be defined in $Q$ as
$x^+ = \bigvee_{i>0} x^i$ for any $x\in Q$, where $x^1=x$ and
$x^{i+1} = x\cdot x^i$, and where $\bigvee$ indicates a supremum. A unital quantale is a quantale with a distinguished element $1$ such that $(Q,\cdot,1)$ forms a monoid.

\begin{prp}
\label{prp:lang-quantale}
  The class $(\Langs(\cC),\subseteq,*,\idlang\cC)$ forms a distributive unital
  quantale.
\end{prp}

\begin{proof}
  This follows from Lemmas \ref{l:SAutAssociativity},
  \ref{l:CompleteLattice} and \ref{l:KleeneUnit}.
\end{proof}
The Kleene plus in $\Langs(\cC)$ is thus represented by the plus
in this quantale.

We briefly discuss the functoriality of languages. A d-functor $F:\cD\to\cC$ defines the function
\begin{equation*}
    \Langs(F):\Langs(\cD)\ni L
    \mapsto
    \{[F_*\Gamma] \mid [\Gamma]\in L\}{\downarrow}
    =
    \{\Words(F)(w) \mid w\in L\}{\downarrow}    
    \in \Langs(\cC),
\end{equation*}    
which fails to be a quantale homomorphism in general: if $\Gamma,\Gamma'\in\tro\cD$, $\tgt(\Gamma)\neq\src(\Gamma')$ and $F(\tgt(\Gamma))=F(\src(\Gamma'))$, then \[\Langs(F)(\{[\Gamma]\}\dcl*\{[\Gamma']\}\dcl)=F(\emptyset)=\emptyset\] but $[F_*\Gamma*F_*\Gamma']\in \Langs(F)(\{[\Gamma]]\}\dcl)*\Langs(F)(\{[\Gamma']]\}\dcl)$.
Instead, we obtain the following result.
\begin{lem}
\label{l:LangsFunctoriality}
If a d-functor $F:\cD\to\cC$ is injective on objects,
then $\Langs(F)$ is a quantale homomorphism.
\end{lem}
\mycomment{ 
\begin{lem}
\label{l:LangsFunctoriality}
For every d-functor $F:\cD\to\cC$ that is injective on objects,
the assignment 
\[
    \Langs(F):\Langs(\cD)\ni L
    \mapsto
    \{[F_*\Gamma] \mid [\Gamma]\in L\}{\downarrow}
    \in \Langs(\cC)
\]
    defines a quantale homomorphism.
\end{lem}
}
\begin{proof}
For $L,L'\in\Langs(\cD)$ we wish to show $\Langs(F)(L*L')=\Langs(F)(L)*\Langs(F)(L')$. First, suppose $[\Gamma]\in \Langs(F)(L*L')$, which is the case if and only if $\Gamma\subsu F_*\Theta$ for some $[\Theta]\in L*L'$. But $[\Theta]\in L*L'$ if and only if there are $[\Delta]\in L$ and $[\Delta']\in L'$ such that $\Theta\subsu \Delta*\Delta'$. 
 Therefore,
    \[
    [\Gamma]\subsu [F_*\Theta] \subsu [F_*(\Delta*\Delta')]
    = [F_*\Delta] * [F_*\Delta']\in \Langs(F)(L)*\Langs(F)(L'),
    \]
using Lemma \ref{l:BasicPropertiesOfDirectImages} in the second and third step, and $[\Gamma]\in \Langs(F)(L)*\Langs(F)(L')$ follows from down-closure. 

Second, suppose $[\Gamma]\in \Langs(F)(L)*\Langs(F)(L')$. Then there exist $[\Lambda]\in \Langs(F)(L)$, $[\Lambda']\in \Langs(F)(L')$, $[\Delta]\in L$ and $[\Delta']\in L'$ such that $\tgt(\Lambda)=\src(\Lambda')$, $\Gamma\subsu\Lambda*\Lambda'$, $\Lambda\subsu F_*\Delta$ and $\Lambda'\subsu F_*\Delta'$. By injectivity of $F$ on objects we obtain $\tgt(\Delta)=\src(\Delta')$. Therefore, by Lemma \ref{l:BasicPropertiesOfDirectImages},
    \[
    [\Gamma]\subsu [\Lambda*\Lambda']\subsu [F_*\Delta]*[F_*\Delta']= F_*([\Delta*\Delta'])\in \Langs(F)(L*L')
    \]
 and $[\Gamma]\in \Langs(F)(L*L')$ follows  from down-closure. 

 It remains to check $\Langs(F)(\bigcup_{i\in I} L_i) = \bigcup_{i\in I}\Langs(F)(L)$ for a family of $\cD$-languages $\{L_i\}_{i\in I}$. This
    is straightforward.
\end{proof}

The unit $\idlang\cC$ is not preserved by $\Langs(F)$ unless $F$ is surjective on objects.

We formulate an easy fact:
\begin{lem}
\label{l:IsoWord}
    Let $F:\cD\to\cC$ be a functor. If $\Words(F):\Words(\cD)\to\Words(\cC)$ is an isomorphism of categories, then $\Langs(F):\Langs(\cD)\to\Langs(\cC)$ is an isomorphism of quantales. 
\end{lem}

\section{Rational languages over d-categories}
\label{se:ratlang}

We now define rational languages and then regular ones.  The unit
language $\idlang{\cC}$ turns out to be non-regular in general,
for reasons explained below, hence we need \emph{non-unital}
algebras to capture them.

Recall that a Kleene algebra is an additively idempotent semiring
$(K,+,0,\cdot,1)$ equipped with a Kleene star $K\to K$ axiomatised as
a fixpoint, as arbitrary suprema are not available in semirings. In
a non-unital Kleene algebra, the $1$ is forgotten and the star 
replaced by the \emph{Kleene plus} $(-)^+:K\to K$, which satisfies
$x + x^+\cdot x^+ \le x^+$ and
$x + y \cdot y\le y\Longrightarrow x^+ \le y$
for all $x,y\in K$. A non-unital Kleene-algebra homomorphism preserves $0$, $+$, $\cdot$ and $(-)^+$.

\begin{df}
  A language is \emph{elementary} if it is of the form $\{w\}\dcl$ for some
  $w\in\Words(\cC)$. The class
  $\Rat(\cC)\subseteq \Langs(\cC)$ of \emph{rational} languages is generated by the elementary languages and $\emptyset$ using $+$, $*$ and $(-)^+$.
\end{df}

The proof of the following fact is routine given Proposition~\ref{prp:lang-quantale}.

\begin{prp}
  The class $(\Rat(\cC),+,\emptyset,*,(-)^+)$ forms a non-unital
  Kleene algebra.
\end{prp}
In the following, we compare elementary and regular languages over different categories. Fix a d-functor $F:\cD\to\cC$.
\begin{lem}
\label{l:ImageOfAtomic}
    If $\Delta\in\tro\cD$, then $\Langs(F)(\{[\Delta]\}\dcl)=\{[F_*\Delta]\}\dcl$.
\end{lem}
\begin{proof}
    If $[\Gamma]\in \Langs(F)(\{\Delta\}{\downarrow})$, then $\Gamma\subsu F_*\Delta'$ for some $\Delta'\in\tro\cD$ such that $\Delta'\subsu\Delta$.
    Thus, $\Gamma\subsu F_*\Delta'\subsu F_*\Delta$ by Lemma \ref{l:BasicPropertiesOfDirectImages} and therefore $\Langs(F)(\{[\Delta]\}\dcl)\subseteq \{[F_*\Delta]\}\dcl$. The inverse inclusion is clear.
\end{proof}
\begin{lem}~
\label{l:RatFunctoriality}
\begin{enumerate}
    \item Assume that the functor $F$ is injective on objects. If $L\in\Rat(\cD)$, then $\Langs(F)(L)\in\Rat(\cC)$, and $\Langs(F)$ restricts to a non-unital Kleene algebra homomorphism $\Rat(F):\Rat(\cD)\to\Rat(\cC)$.
    \item If the functor $\Words(F)$ is an isomorphism of categories, then $\Rat(F)$ is an isomorphism of non-unital Kleene algebras.
\end{enumerate}
    
\end{lem}
\begin{proof}
    Lemma \ref{l:ImageOfAtomic} shows that $\Langs(F)$ preserves elementary languages. The conclusion then follows from Lemma \ref{l:LangsFunctoriality}. The second claim follows from Lemma \ref{l:IsoWord}.
\end{proof}

\mycomment{
\begin{lem}
    Assume that for every $\Gamma\in\tro\cC$ there exists $\Delta\in\tro\cD$ such that $\Gamma\simeq F_*\Delta$. Then the functor $\Words(F)$ and homomorphisms $\Langs(F)$ and $\Rat(F)$ are surjective.
\end{lem}
\begin{proof}
    The assumption is equivalent to the surjectivity of $\Words(F)$.
    This implies that every elementary $\cC$-language belong to the image of $\Rat(F)$ (and hence of to the image of $\Langs(F)$).
    But the set of elementary languages generates both the quantale $\Langs(\cC)$ and the non-unital Kleene algebra $\Rat(\cC)$. 
\end{proof}
}
\mycomment{
\textcolor{brown}{
\begin{lem}
    Assume that $F:\cD\to\cC$ is a d-functor such that:
    \begin{itemize}
        \item Every $\cC$-track object is isomorphic to $F_*\Delta$ for $\Delta\in\tro\cD$,
        \item If $F_*\Delta\subsu F_*\Delta'$, then $\Delta\subsu\Delta'$.
    \end{itemize}
    Then $\Rat(F):\Rat(\cD)\to\Rat(\cC)$ is an isomorphism.
\end{lem}
\begin{proof}
\end{proof}
}
\textcolor{red}{Lemma above is never used. Proof is missing.}
}

The \emph{source} and \emph{target} of a $\cC$-language $L$ are defined as
\[
  \src(L)=\{\src(w)\mid w\in L\}\subseteq \Ob(\cC)
  \quad \text{and}\quad
  \tgt(L)=\{\tgt(w)\mid w\in L\}\subseteq \Ob(\cC).
\]
A $\cC$-language is \emph{simple} if its source and target each have precisely one element.

Every $L\in\Langs(\cC)$ can be partitioned into a disjoint union of simple languages,
\begin{equation}
  \label{e:LangDecomp}
  L=\bigsqcup_{U\in \src(L)} \bigsqcup_{V\in \tgt(L)} {}_UL_V,
\end{equation}
where $\ilo ULV=\{w\in L\mid\src(w)=U,\; \tgt(w)=V\}$. Note that
\begin{equation}
   \label{e:Localization}
  \ilo ULV
  =\{[\idtrack{U}]\}\dcl *L * \{[\idtrack{V}]\}\dcl.
  \end{equation}

\begin{exa}
  In the $\square$-language $L$ from Example \ref{x:PrecubeLanguage}, $\src(L)=\{[0]\}$ and $\tgt(L)=\{[0],[1]\}$.  The
  partition \eqref{e:LangDecomp} becomes $L=\ilo {[0]}L{[1]}\sqcup \ilo {[0]}L{[1]}$, where
  \begin{equation*}
    \ilo {[0]}L{[1]}=
    \left\{
      \vcenter{\hbox{%
          \begin{tikzpicture}[scale=.7, -, transform shape]
            \filldraw[color=black!10!white] (0,0)--(1,0)--(1,1)--(0,1)--(0,0);
            \node[state] (00) at (0,0) {};
            \node[state] (10) at (1,0) {};
            \node[state] (01) at (0,1) {};
            \node[state] (11) at (1,1) {};
            \path (00) edge (10);
            \path (01) edge node[above] {$\top$} (11);
            \path (00) edge (01);
            \path (10) edge (11);
            \node[below] at (00) {$\bot$};
          \end{tikzpicture}}},
      \vcenter{\hbox{%
          \begin{tikzpicture}[scale=.7, -, transform shape]
            \node[state] (00) at (0,0) {};
            \node[state] (01) at (0,1) {};
            \node[state] (11) at (1,1) {};
            \path (01) edge node[above] {$\top$} (11);
            \path (00) edge (01);
            \node[below] at (00) {$\bot$};
          \end{tikzpicture}}}
      \right\},\qquad
    \ilo {[0]}L{[0]}=
    \left\{
      \vcenter{\hbox{%
          \begin{tikzpicture}[scale=.7, -, transform shape]
            \filldraw[color=black!10!white] (0,0)--(1,0)--(1,1)--(0,1)--(0,0);
            \node[state] (00) at (0,0) {};
            \node[state] (10) at (1,0) {};
            \node[state] (01) at (0,1) {};
            \node[state] (11) at (1,1) {};
            \node[state] (21) at (2,1) {};
            \path (00) edge (10);
            \path (01) edge (11);
            \path (11) edge (21);
            \path (00) edge (01);
            \path (10) edge (11);
            \node[below] at (00) {$\bot$};
            \node[above] at (21) {$\top$};
          \end{tikzpicture}}},
      \vcenter{\hbox{%
          \begin{tikzpicture}[scale=.7, -, transform shape]
            \node[state] (00) at (0,0) {};
            \node[state] (10) at (1,0) {};
            \node[state] (11) at (1,1) {};
            \node[state] (21) at (2,1) {};
            \path (00) edge (10);
            \path (11) edge (21);
            \path (10) edge (11);
            \node[below] at (00) {$\bot$};
            \node[above] at (21) {$\top$};
          \end{tikzpicture}}}
    \right\}.
  \end{equation*}

\end{exa}	

\begin{lem}
\label{l:RatSrcTgt}
	Let $L,M\in\Langs(\cC)$. Then
	\begin{enumerate}
	\item 
		$\src(L\cup M)=\src(L)\cup\src(M)$ and $\tgt(L\cup M)=\tgt(L)\cup\tgt(M)$,
	\item
		$\src(L*M)\subseteq \src(L)$ and $\tgt(L*M)\subseteq \tgt(M)$,
	\item
		$\src(L^+)=\src(L)$ and $\tgt(L^+)=\tgt(L)$.
	\end{enumerate}
\end{lem}
\begin{proof}
    This follows immediately from the definitions.
\end{proof}

\begin{lem}
  \label{l:RatSrcTgtFinite}
  A language $L\in\Langs(\cC)$ is rational if and only if
  \begin{enumerate}
  \item the sets $\src(L)$ and $\tgt(L)$ are finite, and
  \item ${}_UL_V$ is rational for all $U\in\src(L)$, $V\in\tgt(L)$.
  \end{enumerate}	
\end{lem}

\begin{proof}
  Suppose $L$ is rational.  By Lemma \ref{l:RatSrcTgt}, the
  rational operations preserve finiteness of sources and targets,
  which implies (1), while (2) follows from
  \eqref{e:Localization}.  Conversely, if (1) and (2) hold, then
  rationality of $L$ follows from \eqref{e:LangDecomp}.
\end{proof}

\begin{cor}
	$\idlang\cC$ is a rational language if and only if the set $\Ob(\cC)$ is finite.
\end{cor}

\begin{exa}
  For the index category defining $\cG$-automata, $\Ob(\cG)=\Sigma \cup \{\emptyset\}$ is finite, which
  explains why the algebra of standard rational languages includes the
  Kleene star.  For $\oldsq$-automata, $\Ob(\oldsq)$ is not finite, and
  the algebra of rational languages of HDA needs to use the Kleene
  plus \cite{FahrenbergJSZ24}.
\end{exa}

\section{Regular languages over d-categories}
\label{se:reglang}

Standard regular languages are recognised by finite automata, yet the
notion of finiteness must be adapted for presheaf automata.  Let $\cC$
be a d-category.
\begin{df}
  A $\cC$-presheaf $X$ is \emph{finitely generated} if there is a
  functor $G:\cE\to\cC$ from some finite category $\cE$ such that
  $X\simeq \widehat{G}$.  A $\cC$-automaton has \emph{finite type} if it has finitely many start and accept cells
  and the underlying presheaf is finitely generated.
\end{df}

\begin{df}
   A $\cC$-language $L$ is \emph{regular} if $L=\Lang(X)$ for some
   finite-type $\cC$-automaton $X$.  We write $\Reg(\cC)$ for the set
   of all regular languages in $\Langs(\cC)$.
\end{df}

A criterion similar to Lemma \ref{l:RatSrcTgtFinite} holds for regular languages.

\begin{lem}
  \label{l:RegSrcTgtFinite}
  A language $L\in\Langs(\cC)$ is regular if and only if
  \begin{enumerate}
  \item the sets $\src(L)$ and $\tgt(L)$ are finite, and
  \item ${}_UL_V$ is regular for all $U\in\src(L)$ and $V\in\tgt(L)$.
  \end{enumerate}	
\end{lem}
\begin{proof}
  Suppose $L$ is regular.  Then $\Lang(X)=L$ for a $\cC$-automaton $X$ having finite type. If $U\in\src(L)$ ($V\in\tgt(L)$),
  then $X$ has a start (accept) cell of type $U$ ($V$), which implies
  (1).  For $U,V\in\cC$, let $\ilo UXV$ be the automaton with the same
  underlying presheaf as $X$ and with
  $\bot_{\ilo UXV}=\bot_X\cap X[U]$ and
  $\top_{\ilo UXV}=\top_X\cap X[V]$.  Then $\ilo UXV$ has finite type
  and $\Lang(\ilo UXV)=\ilo ULV$, which implies (2).

  For the converse direction, fix a language $L\in\Langs(\cC)$ with
  $\src(L)$ and $\tgt(L)$ finite. For all $U\in\src(L)$,
  $V\in\tgt(L)$, let $X(U,V)$ be automata having finite type such that $\Lang(X(U,V))=\ilo ULV$.  Then, by Lemma
  \ref{l:LanguageOfCoproduct},
  \[
    \Lang\left(\coprod_{U\in \src(L)}			\coprod_{V\in\tgt(L)} X(U,V)\right)
    =
    \bigcup_{U\in\src(L)}\bigcup_{V\in\tgt(L)} X(U,V)
    =
    L,
  \]
  and thus $\coprod_{U}\coprod_{V}X(U,V)$ has finite type as a finite
  coproduct of automata having finite type.  Thus $L$ is regular.
\end{proof}

\begin{df}
  A presheaf is \emph{finite} if its category of elements is.  An
  automaton is \emph{finite} if its underlying presheaf is finite (this implies that it has finitely many start and accept cells).
\end{df}

Finiteness is not functorial,
because a natural equivalences between d-categories need not preserve finiteness.
Under suitable assumptions on the index category
$\cC$, finite and finite-type automata coincide.

\begin{lem}
  Suppose all representable $\cC$-presheaves are finite. Then
  \begin{enumerate}
  \item a
    $\cC$-presheaf is finite if and only if it is finitely generated;
  \item a $\cC$-automaton is finite if and only if it has finite
    type.
  \end{enumerate}
  
\end{lem}

\begin{proof}
  Let $X$ be a presheaf. If $X$ is finite, then it is finitely
  generated by $\pi_X$ by Lemma \ref{l:Density}.  Conversely, if $X$
  if finitely generated, that is, $X\simeq \widehat{G}$ for some
  $G:\cE\to\cC$ with $\cE$ finite, then the structural map
  \[
    \bigsqcup_{e\in \cE} \widehat{G(e)}
    \to
    \colim_{e\in\cE} \widehat{G(e)}
    =
    \widehat{G}
    \simeq X
  \]
  is a surjection from a finite presheaf and thus $X$ is finite.
\end{proof}

In the d-categories $\cG(\Sigma)$ and $\square$, all representable
presheaves are finite. 
Hence
all standard automata and $\square$-automata having finite type are finite.
See Sections
\ref{s:VASS} - \ref{s:HDAC} for examples of infinite representable presheaves.

\begin{rem}
  Kleene's theorem for standard automata (or $\cG$-automata) states
  that rational and regular languages coincide. It has
  recently been generalised to higher-dimensional
  automata~\cite{FahrenbergJSZ24}, hence in particular
  $\oldsq$-automata.  In Section~\ref{s:Memory} we present a d-category in which Kleene's theorem fails. Whether $\Rat(\cC)=\Reg(\cC)$ holds
  for other classes of $\cC$-automata remains to be explored.
\end{rem}

\section{Open maps}
\label{se:openmaps}

Simulations and bisimulations are standard notions for comparing
automata. Open maps~\cite{Joyal-Winskel-Nielsen_Bisimulation} provide
a categorical foundation for these concepts. Here we adapt them to
presheaf automata.  Let $\cC$ be a d-category.

\begin{df}
\label{d:OpenMap}
A $\cC$-presheaf map $f:Y\to X$ is \emph{future open} if for every
formorphism $\varphi:V\to U$ in $\cC$, every diagram of solid arrows
\begin{equation}
  \begin{tikzcd}
    \widehat{V} \arrow[d,swap,"\varphi_*"] \arrow[r,"y"] & Y \arrow[d,"f"] \\
    \widehat{U}\arrow[r, swap, "x"] \arrow[ur,dotted,"\overline{y}"] & X
  \end{tikzcd}
\end{equation}	
has a filler $\overline{y}$.  It is \emph{past open} if for every
backmorphism $\varphi$ such a filler exists.
\end{df}

We present two criteria for maps to be future or past open. Both follow immediately from the Yoneda lemma.
\begin{lem}
	A $\cC$-presheaf map $f$
	is a future (past) open map if and only if
	for all $\varphi\in\forM\cC(V,U)$
	($\varphi\in\bckM\cC(V,U)$),
	 $y\in Y[V]$ and $x\in X[U]$
	 such that $X[\varphi](x)=f(y)$
	 there exists $\overline{y}\in Y[U]$
	 such that $Y[\varphi](\overline{y})=y$
	 and $f(\overline{y})=x$.
\end{lem}

\begin{lem}
	A $\cC$-presheaf map $f$
	is future (past) open if each diagram of the form (A) (of the
        form (B)) 
	in $\dCat$ has a filler:
\begin{equation*}
  (A)\;
  \begin{tikzcd}
    \bot_\upstep \arrow[d,left hook->] \arrow[r,"y"] & \Cell(Y) \arrow[d,"\Cell(f)"] \\
    \upstep\arrow[r, swap, "\phi"] \arrow[ur,dotted,->,"\psi"] & \Cell(X)
  \end{tikzcd}
  \qquad\qquad
  (B)\;
  \begin{tikzcd}
    \top_\downstep \arrow[d,left hook->] \arrow[r,"y"] & \Cell(Y) \arrow[d,"\Cell(f)"] \\
    \downstep\arrow[r, swap, "\phi"] \arrow[ur,dotted,->,"\psi"] & \Cell(X)
  \end{tikzcd}
\end{equation*}

\end{lem}

We focus on future open maps in what follows; results for past open
maps are similar.

\begin{lem}
  \label{l:OpenCatCrit}
  Let $f\in \psh\cC(Y,X)$.
  The following conditions are equivalent:
  \begin{enumerate}
  \item
    the map $f$ is future open,
   \item
   for every linear category $\cI$, every diagram (C) in $\dCat$ has a filler,
  \item
for every track object $\Gamma$, every diagram (D) in $\psh\cC$ has a filler,
  \item
for all track objects $\Gamma$, $\Delta$ with $\tgt(\Gamma)=\src(\Delta)$, every diagram (E) in $\psh\cC$ has a filler.
  \end{enumerate}
\begin{equation*}
(C)
    \begin{tikzcd}[ampersand replacement=\&]
      \bot_\cI \arrow[d,left hook->] \arrow[r,"y"] \& \Cell(Y) \arrow[d,"\Cell(f)"] \\
      \cI\arrow[r, swap, "\alpha"] \arrow[ur,dotted,->,"\beta"] \& \Cell(X)
    \end{tikzcd}
    \qquad\quad
(D)
    \begin{tikzcd}[ampersand replacement=\&]
      \widehat{\src(\Gamma)} \arrow[d,left hook->] \arrow[r,"y"] \& Y \arrow[d,"f"] \\
      \Gamma\arrow[r, swap, "{\boldsymbol\alpha}"] \arrow[ur,dotted,->,"{\boldsymbol\beta}"] \& X
    \end{tikzcd}
    \qquad\quad
  (E)
    \begin{tikzcd}[ampersand replacement=\&]
      \Gamma \arrow[d,left hook->] \arrow[r,"\boldsymbol\gamma"] \& Y \arrow[d,"f"] \\
      \Gamma*\Delta\arrow[r, swap, "{\boldsymbol\alpha}"] \arrow[ur,dotted,->,"{\boldsymbol\beta}"] \& X
    \end{tikzcd}
  \end{equation*}
\end{lem}
\begin{proof}
  (2)$\iff$(3) follows from the Yoneda lemma;
  (4)$\implies$(3)$\implies$(1) is obvious.  It remains to prove
  (1)$\implies$(4).

  Suppose $f$ is future open.  We show that every diagram (E) has a
  filler by induction on the length of decompositions of $\Delta$ into
  elementary tracks.
  
  If $\Delta$ is an identity track object, then
  $\Gamma\cong\Gamma*\Delta$ and there is nothing to prove.  If
  $\Delta\simeq ( \widehat{U}\overset{\id}\longrightarrow\widehat{U}
  \overset{\widehat{\psi}}\longleftarrow \widehat{V})$ for $\psi\in\bckM\cC(V,U)$, then the inclusion
  $\Gamma\to \Gamma*\Delta$ is an isomorphism of presheaves and
  again we are done.
  
  If $\Delta\simeq   (\widehat{V}\overset{\widehat\phi}\longrightarrow\widehat{U}
  \overset{\id}\longleftarrow \widehat{U})
$ for
  $\varphi\in\forM\cC(V,U)$, then we consider the diagram
  \[
    \begin{tikzcd}[row sep=large]
      \widehat{V} \arrow[d,"{\varphi_*}" swap] \arrow[r,"\top"]
      &
      \Gamma \arrow[d,left hook->] \arrow[r,"\boldsymbol\gamma"]
      &
      Y \arrow[d,"f"]
      \\
      \widehat{U}\arrow[r,right hook->] \arrow[urr,dotted,->,"\boldsymbol\delta",pos=0.3]
      &
      \Gamma*\Delta\arrow[r, swap, "\boldsymbol\alpha"] \arrow[ur,dotted,->,"\boldsymbol\beta" swap]
      &
      X
    \end{tikzcd}
  \]
  of solid arrows. The left-hand square is a pushout diagram defining
  $\Gamma*\Delta$.  By Definition \ref{d:OpenMap}, it has a filler
  $\boldsymbol\delta$, since $f$ is future open and $\varphi$ a
  formorphism.  Further, the maps $\boldsymbol\gamma$ and
  $\boldsymbol\delta$ induce a map $\boldsymbol\beta$ which is a
  required filler.
  
  If $\Delta$ is not elementary, then there exists a presentation $\Delta=\Delta'*\Delta''$
  with $\Delta'$, $\Delta''$ of smaller lengths:
  \[
    \begin{tikzcd}[column sep=large]
      \Gamma
      \arrow[d,"i" swap]
      \arrow[r,"\boldsymbol\gamma"]
      &
      Y
      \arrow[dd,"f"]
      \\
      \Gamma*\Delta'
      \arrow[d,"j" swap]
      \arrow[ur,dotted,"\boldsymbol\delta"]
      &
      \\
      \Gamma*\Delta'*\Delta''
      \arrow[r, swap, "\boldsymbol\alpha"]	  
      \arrow[uur,dotted,"\boldsymbol\beta" swap]
      &
      X
    \end{tikzcd}
  \]
  Using the induction hypothesis twice produces a map
  $\boldsymbol\delta$ and then a map $\boldsymbol\beta$.
\end{proof}

\begin{prp}
  Let $f:Y\to X$ be a future open map of $\cC$-automata such that
  $\bot_X\subseteq f(\bot_Y)$ and $\top_Y=f^{-1}(\top_X)$.  Then
  $\Lang(Y)=\Lang(X)$.
\end{prp}
\begin{proof}
  First, $\Lang(Y)\subseteq \Lang(X)$ follows from Lemma
  \ref{l:LangFun}.  Conversely, suppose $[\Gamma]\in\Lang(X)$. Then
  there is an accepting track $\boldsymbol\alpha:\Gamma\to X$.  By
  hypothesis, there is an element $y\in \bot_Y$ such that
  $f(y)=\src(\boldsymbol\alpha)$.  Let $\boldsymbol\beta:\Gamma\to Y$
  be a filler for the diagram (D) of Lemma \ref{l:OpenCatCrit}.
  Clearly $\src(\boldsymbol\beta)=y\in\bot_Y$ and
  $\tgt(\boldsymbol\beta)\in \top_Y$ since
  $f(\tgt(\boldsymbol\beta))=\tgt(\boldsymbol\alpha)\in\top_X$.  Thus,
  $\boldsymbol\beta$ is accepting and $[\Gamma]\in\Lang(Y)$.
\end{proof}

\section{Geometric models of concurrency}
\label{s:HDA}

\def\CSet{\mathbf{CSet}}
\def\CList{\mathbf{CList}}
\def\SCSet{\mathbf{SCSet}}
\def\SCList{\mathbf{SCList}}

As a proof of concept we now review several geometric models for
concurrency \cite{FJSZ_HDALang, FahrenbergJSZ24, FGHMR_DAT,
  Grandis-Mauri_Cubical, vanGlabbeek_Expressiveness} and represent
them as presheaf automata.  We present most  relationships without proofs, as they are rather obvious, but tedious to check in detail. The fundamental notion is that of consets
(concurrency sets) and concset maps, which is motivated by the
conclists and conclist maps for HDA 
\cite{FJSZ_HDALang,FahrenbergJSZ24}. The resulting categories
replace the precubical category $\oldsq$ from
Section~\ref{s:PrecubicalSets} as index categories for
HDA and similar models. In particular we now
consider variants with labels, which is standard for concurrency.

We fix an infinite set and call the finite subsets of this set
\emph{concsets} (to enforce smallness of the categories below).

\begin{df}
  A \emph{concset map} from a concset $V$ to a concset $U$ is a pair
  $(f,\varepsilon)$ where $f:V\to U$ is an injection and $\varepsilon:U\to \{0,*,1\}$ satisfies
  $\varepsilon^{-1}(*)=f(V)$.  The composition of conset maps
  $(g,\eta):W\to V$ and $(f,\varepsilon):V\to U$ is the concset map
  $(gf,\theta):W\to U$ with
  \begin{equation*}
    \label{e:ConcsetComposition}
    \theta(u)=
    \begin{cases}
      \varepsilon(u) & \text{for } u\not\in f(V),\\
      \eta(f^{-1}(u)) & \text{for } u\in f(V).
    \end{cases}
  \end{equation*}
  A concset map $(f,\varepsilon):V\to U$ is a formorphism if
  $\varepsilon(u)\neq 1$ for all $u\in U$, and a backmorphism
  if $\varepsilon(u)\neq 0$ for all $u\in U$.

  A \emph{pointed concset map} $(f,\varepsilon):(V,v_0)\to (U,u_0)$
  between pointed concsets $(U,u_0),(V,v_0)$ is defined as above, but
  $f$ must preserve basepoints ($f(v_0)=u_0$) and may map many
  elements to the basepoint (if $f(v)=f(v')=u$, then $v=v'$ or
  $u=u_0$).  The above formula for concset composition remains valid.
\end{df}

\begin{rem}
  There is a bijection between pointed concset maps
  $(V,v_0)\to (U,u_0)$ and partial concset maps $V\to U$, meaning that $f$ is not defined on the whole domain.
\end{rem}

Composition of concset maps (pointed or unpointed) is associative.
The identity on $U$ or $(U,u_0)$ is
$(\id_U,U\ni u \mapsto *\in\Omega)$. This allows defining several
d-categories.

\begin{enumerate}
\item the concset category $\CSet$ of concsets and concset maps;
\item
	the conclist category $\CList$
	of concsets with a strict total order $\intord$ (\emph{conclists})
	and order-preserving concset maps (\emph{conclist maps});
\item
	the pointed concset category $\CSet_*$
	of pointed concsets and basepoint-preserving concset maps;
\item
	the pointed conclist category $\CList_*$
	of pointed conclists, whose
	 objects are pointed concsets $(U,u_0)$
	with a total order $\intord$ on $U\setminus\{u_0\}$,
	and morphisms are basepoint-preserving concset maps
	$(f,\varepsilon):(V,v_0)\to(U,u_0)$ such that $f$ restricted
	to $f^{-1}(U\setminus\{u_0\})$ preserves $\intord$;
\item
	the $\Sigma$-labelled concset category $\CSet(\Sigma)$
	over the alphabet $\Sigma$,
	of $\Sigma$-labelled concsets,
	which are pairs $(U,\lambda:U\to \Sigma)$,
	and label-preserving concset maps;
\item
	the $\Sigma$-labelled conclist category $\CList(\Sigma)$
	of labelled conclists and order-and-label-preserving concset maps.
\end{enumerate}
Presheaves and automata on these d-categories coincide  with the following existing models for concurrency:
\begin{enumerate}
\item The conclist category $\CList$ is equivalent to the category
  $\oldsq$ from Definition \ref{d:PrecubeCategory}. The
  functor $I:\oldsq\to\CList$ given by
  $I([n])=(1\intord \dotsm \intord n)$,
  $I(d^\alpha_i)=(\delta_i,\varepsilon^\alpha_i)$, where
\begin{equation}
\label{e:DeltaEpsilon}
  \delta_i(k)
  =
  \begin{cases}
    k & \text{for }k<i,\\
    k+1 & \text{for }k\geq i,
  \end{cases}
  \qquad
  \varepsilon^\alpha_i(m)
  =\begin{cases}
    \alpha & \text{for }m=i,\\
    * & \text{for }m\neq i,
  \end{cases}
\end{equation}
gives rise to an equivalence of categories.  $\CList$-presheaves are
thus equivalent to the precubical sets of
\cite{FJSZ_HDALang,FGHMR_DAT,vanGlabbeek_Expressiveness} 
(called cubical sets in the last reference).
\item
\label{i:CListStar}
The pointed conclist category $\CList_*$ is equivalent to the cubical
category $\square$ of \cite{Goubault-Mimram_Relationship} and the
restricted cubical site $\mathbb{I}$ of \cite{Grandis-Mauri_Cubical}.
\item The concset category $\CSet$ and the pointed concset category
  $\CSet_*$ are equivalent to the symmetric precubical category and
  the symmetric cubical category of
  \cite{Goubault-Mimram_Relationship}, respectively.
\item The $\Sigma$-labelled conclist category $\CList(\Sigma)$ is
  equivalent to the precubical category $\square$ of
  \cite{FahrenbergJSZ24}.  $\CList(\Sigma)$-automata are thus
  equivalent to the higher-dimensional automata studied in
  \cite{FJSZ_HDALang,FahrenbergJSZ24,vanGlabbeek_Expressiveness}.
\item Presheaves on the labelled concset category $\CSet(\Sigma)$
  are equivalent to the labelled symmetric precubical sets of
  \cite{Goubault-Mimram_Relationship} and closely related to the
  cubical site $\mathbb{J}$ of \cite{Grandis-Mauri_Cubical}.
\end{enumerate}

We outline the equivalences claimed in item \ref{i:CListStar} in
this list as an example. 
The proofs of items 3--5 are not entirely straightforward, but similar; details are left for future work.

The \emph{cubical category} $\mathbb{I}$ \cite{Grandis-Mauri_Cubical}
is the d-category defined by the following data:
\begin{itemize}
\item
  objects are ordinals $[n]$ for $n\geq 0$,
\item morphisms are generated by
  $d^0_i,d^1_i\in \mathbb{I}([n-1],[n])$ and
  $e_i\in\mathbb{I}([n+1],[n])$ for $1\leq i\leq n$, and
  relations
  \[
    d^\varepsilon_{j}\circ d^\eta_{i}=d^\eta_{i}\circ d^\varepsilon_{j-1}
    \quad(j<i),
    \qquad
    e_{j}\circ e_i=e_{i-1}\circ e_j
    \quad(j<i),
  \]
  \[
    e_i\circ d_j
    =
    \begin{cases}
      d_j\circ e_{i-1} & \text{for }j<i,\\
      \id & \text{for }j=i,\\
      d_{j-1}\circ e_i & \text{for }j>i,
    \end{cases}
  \]
 
\item
  $\forM{\mathbb{I}}$ is generated by the morphisms $d^0_i$ and $e_i$,
\item
  $\bckM{\mathbb{I}}$ is generated by the morphisms $d^1_i$ and $e_i$.
\end{itemize}
We define a functor $F:\mathbb{I}\to \CList_*$ as
\[
  F([n])=(\{0,1\intord\dotsm\intord n\},0)
\]
on objects and $F(d^\alpha_i:[n-1]\to[n])=(\delta_i,\varepsilon^\alpha_i)$, $F(e_i:[n+1]\to[n])=(g_i,\zeta_i)$ on morphisms, where
\begin{gather*}
  g_i(k)
  =
  \begin{cases}
    k & \text{for }k<i,\\
    k-1 & \text{for }k>i,\\
    0 & \text{for }k=i,
  \end{cases}
  \qquad
  \zeta_i(m)=*.
\end{gather*}
and $\delta_i$ and $\varepsilon^\alpha_i$ are is in \eqref{e:DeltaEpsilon}.
It can be shown that $F$ gives rise to an equivalence of d-categories.

\begin{rem}
  In the terminology of \cite{Grandis-Mauri_Cubical}, the conclist
  category $\CList$ contains only face
  morphisms.  Passing from conclists to
  consets introduces symmetries. Passing from unpointed to
  pointed conclists or concsets adds degeneracies to the respective
  categories. 
\end{rem}
\begin{rem}
Our general approach allows defining track objects and languages for all categories mentioned above. Languages for $\CList(\Sigma)$, hence implicitly for $\CList$, have been calculated in \cite{FJSZ_HDALang}; those for the other cases are not straightforward and left for future work.
\end{rem}

\section{Vector additions systems with states as presheaf automata}
\label{s:VASS}

\def\vvec{\mathsf{vec}}

We now show that vector addition systems with states (VASSes)
\cite{Hopcroft-Pansion_Reachability} can be realised as presheaf
automata.  Vector additions systems and Petri nets are therefore
subsumed by presheaf automata.

\begin{df}
\label{d:VASS}
 An \emph{$r$-dimensional vector addition system with states}
(\emph{$r$-VASS}) is a pair $(Q,E)$, where
\begin{itemize}
\item
 $Q$ is a finite set of \emph{vertices},
 \item
 $E\subseteq Q\times \Z^r\times Q$
is a finite set of \emph{edges}.
\end{itemize}
We write $e=(\src(e),\vvec(e),\tgt(e))$ for
$e\in E$.  

A \emph{run} of a VASS $(Q,E)$ is a sequence
\begin{equation}
\label{e:VassRun}
	(q_0,\sfv_0)\xrightarrow{e_1}(q_1,\sfv_1)\xrightarrow{e_2}
	\dotsm\xrightarrow{e_n}(q_n,\sfv_n)
\end{equation}
such that $q_k\in Q$,
$\sfv_k\in\N^r$,
$e_k\in E$, and further 
$\src(e_k)=q_{k-1}$, 
$\tgt(e_k)=q_k$
and $\sfv_{k}=\sfv_{k-1}+\mathsf{vec}(e_k)$
for all $k$.
\end{df}

An $r$-VASS is thus essentially a $\Z^r$-labelled digraph $G_{(Q,E)}$
and, therefore, a $\cG(\Z^r)$-presheaf. Yet, such a presentation does not
yield the correct semantics. Instead, we enrich the
underlying d-category.

For a set $A$, we write $\cF(A)$ for the
category with $\Ob(\cF(A))=A$ and $\cF(A)(a,b)=\{\iota_a^b\}$ for all
$a,b\in A$.  ($\cF(A)$ is naturally equivalent to the trivial
category.)  For a vector $\sfu\in\Z^r$, let $\sfu=\sfu^+-\sfu^-$ be the
decomposition into the positive and negative parts
($\sfu^+,\sfu^-\in\N^r$).

Let $\cV_r$ be the d-category with the underlying category
$\cG(\Z^r)\times \cF(\N^r)$ and d-structure given, for all
$\sfu\in\Z^r$ and $\sfv\in\N^r$, by
\begin{align*}
	\forM{\cV_r}
	&=\{(\sigma_{\sfv},\iota_{\sfu^-+\sfv}^\sfv)
	:(\vo,\sfu^-+\sfv)\to(\sfu,\sfv)\}
	\cup\{\text{identities}\},
\\
	\bckM{\cV_r}
	&=\{(\tau_{\sfv},\iota_{\sfu^++\sfv}^\sfv)
	:(\vo,\sfu^++\sfv)\to(\sfu,\sfv)\}
	\cup\{\text{identities}\}.
\end{align*}

There is a correspondence between $r$-VASSes and $\cV_r$-presheaves.
\begin{itemize}
\item
For every $r$-VASS $(Q,E)$ we define the $\cV_r$-presheaf $X=X_{(Q,E)}$
as
\begin{align*}
	X[(\vo ,\sfv)]
	&=
	\{(q,\sfv)\mid q\in Q\},
\\
	X[(\sfu ,\sfv)]
	&=
	\{(e,\sfv)\mid e\in E,\; \vvec(e)=\sfu \}
\end{align*}
on objects and
\begin{align*}
	X[(\id_\vo, \iota_\sfw^\sfv)](q,\sfv)
	&= (q,\sfw),
	&
	X[(\id_\sfu, \iota_\sfw^\sfv)](e,\sfv)
	&=(e,\sfw),
\\
	X[(\sigma_\sfu, \iota_\sfw^\sfv)](e,\sfv)
	&= (\src(q), \sfw),
	&
	X[(\tau_\sfu, \iota_\sfw^\sfv)](e,\sfv)
	&= (\tgt(q), \sfw)
\end{align*}
on morphisms. See Figure \ref{f:ElementsOfVASS} for an example.
\item
For every $\cV_r$-presheaf $X$, the corresponding VASS $(Q_X,E_X)$ is given by
\begin{align*}
Q_X
&=X[(\vo,\sfO)],\\
E_X
&=\bigcup_{\sfu\in\Z^r}
\{
	(
X[(\sigma_\sfu,\iota_{\mathsf{0}}^{\mathsf{0}})](e),
		\sfu,
		X[(\tau_\sfu,\iota_{\mathsf{0}}^{\mathsf{0}})](e)
		)
		\mid
		e\in X[(\sfu,\mathsf{0})]
	\}.
\end{align*}
\end{itemize}

\begin{figure}
{
\begin{center}
\begin{tikzpicture}
\begin{scope}[shift={(2,-2)}]
	\node[state, minimum size=10pt] (x) at (0,0) {$x$};
	\node[below left] at (x) {$\bot$};
	\node[state, minimum size=10pt] (y) at (2,0) {$y$};
	\node[state, minimum size=10pt] (z) at (4,0) {$z$};
	\node[above right] at (z) {$\top$};
	\path (x) edge node[above] {$\begin{pmatrix} 2\end{pmatrix}$} node[below]{$a$} (y);
	\path (y) edge node[above] {$\begin{pmatrix} -1\end{pmatrix}$} node[below]{$b$} (z);
\end{scope}
\begin{scope}[shift={(0,0)}, yscale=0.6]
	\node at (0,8) {$\dots$};
	\node at (2,8) {$\dots$};
	\node at (4,8) {$\dots$};
	\node at (6,8) {$\dots$};
	\node at (8,8) {$\dots$};
	\node[state, minimum size=10pt] (x0) at (0,0) {$x,0$};
	\node[state, minimum size=10pt] (a0) at (2,0) {$a,0$};
	\node[state, minimum size=10pt] (y0) at (4,0) {$y,0$};
	\node[state, minimum size=10pt] (b0) at (6,0) {$b,0$};
	\node[state, minimum size=10pt] (z0) at (8,0) {$z,0$};
	\node[state, minimum size=10pt] (x1) at (0,2) {$x,1$}; \node at (-0.6,2) {$\bot$};
	\node[state, minimum size=10pt] (a1) at (2,2) {$a,1$};
	\node[state, minimum size=10pt] (y1) at (4,2) {$y,1$};
	\node[state, minimum size=10pt] (b1) at (6,2) {$b,1$};
	\node[state, minimum size=10pt] (z1) at (8,2) {$z,1$};
	\node[state, minimum size=10pt] (x2) at (0,4) {$x,2$};
	\node[state, minimum size=10pt] (a2) at (2,4) {$a,2$};
	\node[state, minimum size=10pt] (y2) at (4,4) {$y,2$};
	\node[state, minimum size=10pt] (b2) at (6,4) {$b,2$};
	\node[state, minimum size=10pt] (z2) at (8,4) {$z,2$};  \node at (8.6,4) {$\top$};
	\node[state, minimum size=10pt] (x3) at (0,6) {$x,3$};
	\node[state, minimum size=10pt] (a3) at (2,6) {$a,3$};
	\node[state, minimum size=10pt] (y3) at (4,6) {$y,3$};
	\node[state, minimum size=10pt] (b3) at (6,6) {$b,3$};
	\node[state, minimum size=10pt] (z3) at (8,6) {$z,3$};
	\path (x0) edge[thick,color=green!50!black] node[above] {\scriptsize $(\sigma_{2},\iota_0^0)$} (a0);
	\path (x1) edge[thick,color=green!50!black] node[above] {\scriptsize $(\sigma_{2},\iota_1^1)$} (a1);
	\path (x2) edge[thick,color=green!50!black] node[above] {\scriptsize $(\sigma_{2},\iota_2^2)$} (a2);
	\path (x3) edge[thick,color=green!50!black] node[above] {\scriptsize $(\sigma_{2},\iota_3^3)$} (a3);
	\path (y2) edge[thick,color=red!50!black] node[above,rotate=60] {\scriptsize $(\tau_{2},\iota_2^0)$} (a0);
	\path (y3) edge[thick,color=red!50!black] node[above,rotate=60] {\scriptsize $(\tau_{2},\iota_3^1)$} (a1);
	\path (y1) edge[thick,color=green!50!black] node[above,rotate=-40] {\scriptsize $(\sigma_{-1},\iota_1^0)$} (b0);
	\path (y2) edge[thick,color=green!50!black] node[above,rotate=-40] {\scriptsize $(\sigma_{-1},\iota_2^1)$} (b1);
	\path (y3) edge[thick,color=green!50!black] node[above,rotate=-40] {\scriptsize $(\sigma_{-1},\iota_3^2)$} (b2);
	\path (z0) edge[thick,color=red!50!black] node[above] {\scriptsize $(\tau_{-1},\iota_0^0)$} (b0);
	\path (z1) edge[thick,color=red!50!black] node[above] {\scriptsize $(\tau_{-1},\iota_1^1)$} (b1);
	\path (z2) edge[thick,color=red!50!black] node[above] {\scriptsize $(\tau_{-1},\iota_2^2)$} (b2);
	\path (z3) edge[thick,color=red!50!black] node[above] {\scriptsize $(\tau_{-1},\iota_3^3)$} (b3);
\end{scope}
\end{tikzpicture}
\end{center}
}
\caption{
\label{f:ElementsOfVASS}
Example of a VASS and above it the category of elements of the corresponding
$\cV_1$-automaton.
The initial and final configurations of the VASS are $(1)$ and $(2)$, respectively.
 Formorphism are shown in green, backmorphisms in
red. All other morphisms are not shown. 
This $\cV_1$-automaton is a track object.}
\end{figure}

\begin{prp}~
\begin{enumerate}
\item There is a 1-1 correspondence between r-VASSes $(Q,E)$ and $\cV_r$-presheaves $X$ of finite type without ``double'' edges,
	that is, if $e_1,e_2\in X[(\sfu,\sfO)]$
	then $X[(\sigma_\sfu,\iota_\sfO^\sfO)](e_1)
	\neq X[(\sigma_\sfu,\iota_\sfO^\sfO)](e_2)$
	or
	$X[(\tau_\sfu,\iota_\sfO^\sfO)](e_1)
	\neq X[(\tau_\sfu,\iota_\sfO^\sfO)](e_2)$.
    
\item There is also a 1-1 correspondence between runs of $(Q,E)$ and equivalence classes of paths on $X$ that start and terminate at vertices.
    \end{enumerate}
\end{prp}
\begin{proof}
  Checking that the operations presented above are inverses is straightforward. Since $\cV_r$ and $\cG(\Z^r)$ are equivalent categories, the presheaf $X$ has finite type if and only if its restriction to $\cG(\Z^r)\times\{\mathsf{0}\}$ does. The latter condition holds if and only if $Q_X$ and $E_X$ are both finite.

A path on $X$ corresponding to the run \eqref{e:VassRun} is 
    \[
(q_0,\sfv_0)
	\arrO{(\sigma_{\vvec(e_1)}, \iota_{\sfv_0}^{\sfw_1})}
	(e_1,\sfw_1)
	\arrI{(\tau_{\vvec(e_1)},\iota_{\sfv_1}^{\sfw_1})}
	(q_1,\sfv_1)
	\arrO{(\sigma_{\vvec(e_2)}, \iota_{\sfv_1}^{\sfw_2})}
	(\sfu_2,\sfw_2)
	\dotsm    
    \]
    where $\sfw_k=\sfv_{k-1}-\vvec(e_k)^-$, $\sfv_k=\sfw_k+\vvec(e_k)^+$. Again, it is straightforward to check that this correspondence is bijective. 
\end{proof}

We briefly discuss track objects and the category of $\cV_r$-words. Every path in $\cV_r$ that starts and terminates in a vertex is isomorphic a path of the form
	\begin{multline*}
	(\vo,\sfv_0)
	\arrO{(\sigma_{\sfu_1},\iota_{\sfv_0}^{\sfw_1})}
	(\sfu_1,\sfw_1)
	\arrI{(\tau_{\sfu_1},\iota_{\sfv_1}^{\sfw_1})}
	(\vo,\sfv_1)
	\arrO{(\sigma_{\sfu_2},\iota_{\sfv_1}^{\sfw_2})}
	(\sfu_2,\sfw_2)
	\arrI{(\tau_{\sfu_2},\iota_{\sfv_2}^{\sfw_2})}
	\dotsm\\
	\dotsm
	\arrO{(\sigma_{\sfu_n},\iota_{\sfv_{n-1}}^{\sfw_n})}
	(\sfu_n,\sfw_n)
	\arrI{(\tau_{\sfu_n},\iota_{\sfv_n}^{\sfw_n})}
	(\vo,\sfv_n)
	\end{multline*}
	 with $\sfw_k+\sfu_k^+=\sfv_k=\sfw_{k+1}+\sfu_{k+1}^-$,
     $\sfv_k\in \N^r$, $\sfw_k\in\Z^r$.
	The projection functor $\cV_r\to\cG(\Z^r)$
	is an isomorphism of categories -- though not of d-categories. Presheaves on $\cV_r$ are thus equivalent to presheaves on $\cG(\Z^r)$, which are $\Z^r$-labelled graphs.
    Hence, by Lemma \ref{l:TracksOfG}, the set of isomorphism classes of track objects $\Gamma$ in $\cV_r$ with $\src(\Gamma)=(*,\sfv)$ and $\tgt(\Gamma)=(*,\sfv')$ can be identified with sequences $(\sfw_1,\dotsc,\sfw_n)\in(\Z^r)^*$ such that
    \begin{equation}
    \label{e:VASSPositivityCond}
        \sfv+\sfw_1+\dotsm+\sfw_n=\sfv'
        \quad\text{and}\quad
        \sfv+\sfw_1+\dotsm+\sfw_k\geq \mathsf{0}\quad \text{for all $k$}.
    \end{equation}
     As a consequence, we obtain the following.
    
\begin{lem}
    The category $\Words_{\{*\}\times \N^r}(\cV_r)$ has objects $\N^r$ and morphisms
    \[
        \Words_{\{*\}\times \N^r}(\cV_r)(\sfv,\sfv')
        =
        \{(\sfv;\sfw_1,\dotsc,\sfw_n;\sfv')\in \N\times\Z^n\times \N\mid n\geq 0 \text{ and }\eqref{e:VASSPositivityCond}\text{ holds}\}.
    \]
    Composition is given by
    \[
        (\sfv;\sfw_1,\dotsc,\sfw_n;\sfv')*
        (\sfv';\sfw'_1,\dotsc,\sfw'_m;\sfv'')
        =
        (\sfv;\sfw_1,\dotsc,\sfw_n,\sfw'_1,\dotsc,\sfw'_n;\sfv''),
    \]
    and the subsumption order is trivial.
\end{lem}
In the absence of non-trivial subsumptions, $\cV_r$-languages are just sets of $\cV_r$-words. 
Figure \ref{f:ElementsOfVASS} shows an example of a $\cV_1$-track object.

\section{Automata with counters}
\label{s:Memory}

In this section, we present an alternative construction that generalises that of the previous section.  It prepares the definition of higher-dimensional automata with counters in the following section.

We regard a monoid $M$ as a d-category with one single object $\star$ and
$M$ as a set of endomorphisms on $\star$. The only formorphism, and at the
same time backmorphism, is the identity.
\begin{df}
	Let $\cC$ be a d-category.
	A \emph{$\cC$-automaton with counter $M$} 
	is a $(\cC\times M)$-automaton.
\end{df}

$(\cC\times M)$-presheaves are naturally equivalent to functors
$\cC^{op}\to M\mhyphen\Set$ into the category of sets with an action of the monoid $M$ and $M$-equivariant maps.

We realise VASSes as digraphs with counter $\N^r$,
extending the ideas of \cite{Meseguer-Montanari_PetriNetsAreMonoids}.
In the following examples, $\cG=\cG(\{a\})$ is the d-category from Section \ref{s:StdAut}
for a single-letter alphabet.

\begin{exa}
\label{x:VASStoCounter}
	For each VASS $(Q,E)$ we define the $(\cG\times \N^r)$-automaton $X=X_{(Q,E)}$
	with
	\begin{itemize}
	\item
		vertices $X[(\vo,\star)]=Q\times \N^r$, which correspond to configurations on $Q$,
	\item
		edges $X[(a,\star)]=E\times \N^r$,
	\item
		$X[(\id_\vo,\sfw)](q,\sfv)=(q,\sfv+\sfw)$,
	\item
		$X[(\id_a,\sfw)](e,\sfv)=(e,\sfv+\sfw)$,
	\item
		$X[(\sigma_a,\sfw)](e,\sfv)=(\src(e), \sfv+\vvec(e)^-+\sfw)$,
	\item
		$X[(\tau_a,\sfw)](e,\sfv)=(\tgt(e), \sfv+\vvec(e)^++\sfw)$.
	\end{itemize}
Again, there is 1-1 correspondence between runs of $(Q,E)$ and paths on $X$ that start and terminate at vertices. The run \eqref{e:VassRun} translates to
\[
(q_0,\sfv_0)
\arrO{(\sigma_a,\sfO)}
(e_1,\sfw_1)
\arrI{(\tau_a,\sfO)}
(q_1,\sfv_1)
\arrO{(\sigma_a,\sfO)}
\dots
\arrO{(\sigma_a,\sfO)}
(e_n,\sfw_n)
\arrI{(\tau_a,\sfO)}
(q_n,\sfv_n),
\]
where, as above, $\sfw_k=\sfv_{k-1}-\vvec(e_k)^-$, $\sfv_k=\sfw_k+\vvec(e_k)^+$.
Conversely, every path in $X$ with no identity steps has such a form.
\end{exa}

Yet not all $(\cG\times \N^r)$-automata can be realised this way, as the following example shows.

\begin{exa}
	Let $m,n>0$ and let $X$ be a $(\cG\times \N)$-automaton with
	\begin{align*}
		X[(\vo,\star)]
		&=\{x\}\times \{0,\dotsc,n-1\}\cup \{y\}\times \{0,\dotsc,m\},
		\\
		X[(a,\star)]
		&=\{v\}\times \N,
		\\
		X[(\sigma_a,j)](v,k)
		&=(x, (k+j) \mod n),
		\\\
		X[(\tau_a,j)](v,k)
		&=
		\begin{cases}	
			k+j & \text{for } k+j<m,\\
			m & \text{for } k+j \geq m.
		\end{cases}
	\end{align*}
	This can be obtained from an automaton corresponding to a VASS
        with two vertices $x,y$ and a single edge $e$ by identifying
        configurations in $x$ that differ by $n$ and all configurations
        in $y$ with value at least $m$. This automaton does not
        correspond to any VASS, since the set $X[(\vo,\star)]$ is finite.
\end{exa}

Languages of $(\cC\times M)$-automata and $\cC$-automata turn out to be the same.
Let $\Pi:\cC\times M\to\cC$ be the projection on the first factor;
let $I:\cC\to \cC\times M$ be the d-functor given by 
$I(U)=(U,\star)$
for objects $U$ and $I(\varphi)=(\varphi,1_M)$ for morphisms $\varphi$. Obviously, $\Pi\circ I$ is the identity functor on $\cC$.
\begin{lem}~
\label{l:TrackObjectsWithCounters}
\begin{enumerate}
    \item 
	Each track object on $\cC\times M$ is isomorphic to $I_*\Gamma$ for some $\Gamma\in\tro{\cC}$. 
    \item Each track object on $\cC$ is isomorphic to $\Pi_*\Delta$ for some $\Delta\in\tro{\cC\times M}$. 
    \end{enumerate}
\end{lem}
\begin{proof}
	All formorphisms and backmorphisms in $\cC\times M$
	have the form $(\varphi,1_M)$.
	Every path $\omega:\cI\to\cC\times M$
	is thus a composition $\cI\xrightarrow{\eta}\cC\xrightarrow{I}\cC\times M$,
	and every $(\cC\times M)$-track object is isomorphic to
	\[
		\widehat{\omega}
		=
		\colim_{i\in \cI} \widehat{\omega(i)}
		=
		\colim_{i\in \cI} \widehat{(I\eta)(i)}
		=
		\colim_{i\in \cI} I_*\widehat{\eta(i)}
		=
		I_*\colim_{i\in \cI} \widehat{\eta(i)}
		=
		I_*\widehat{\eta},
	\]
    which proves the first statement.
    For $\Gamma\in\tro\cC$, we have $\Pi_*(I_*\Gamma)\simeq \Gamma$, proving the second statement.
\end{proof}
\begin{prp}
\label{p:LanguagesOfCounters}
    For every d-category $\cC$ and monoid $M$, the functors
    \[
        \Words(I):\Words(\cC)\rightleftarrows \Words(\cC\times M):\Words(\Pi)
    \]
    are isomorphisms. They induce isomorphisms of quantales $\Langs(\cC)\simeq \Langs(\cC\times M)$ and of non-unital Kleene algebras $\Rat(\cC)\simeq\Rat(\cC\times M)$.
\end{prp}
\begin{proof}
    $\Words(I)$ and $\Words(\Pi)$ are surjective by Lemma \ref{l:TrackObjectsWithCounters} and $\Words(\Pi)\circ \Words(I)$ is the identity functor. Thus, $\Words(I)\circ\Words(\Pi)$ is also the identity.
    The remaining isomorphisms follow from Lemmas \ref{l:IsoWord} and \ref{l:RatFunctoriality}.
\end{proof}

\begin{exa}
\label{x:Dyck}
Let $\cG=\cG(\{a,b\})$. We construct a $(\cG\times\N)$-automaton of finite type that recognises the Dyck language~\cite[p.\@ 144]{ChomskyS63}, so that this language is $(\cG\times\N)$-regular.
Let $D$ be the following $(\cG\times\N)$-automaton:
  \begin{gather*}
    D[(\vo,\star)]=\{x_k\}_{k\geq 0},
    \qquad
    D[(a,\star)]=\{a_k\}_{k\geq 0},
    \qquad
    D[(b,\star)]=\{b_k\}_{k\geq 0},\\
    D[(\id_\vo,v)](x_k)=x_{k+v},
    \qquad
    D[(\id_a,v)](a_k)=a_{k+v},
    \qquad
    D[(\id_b,v)](x_k)=b_{k+v},
    \\
    D[(\sigma_a,v)](a_k)=D[(\tau_b,v)](b_k)=x_{k+v},
    \\
    D[(\tau_a,v)](a_k)=D[(\sigma_b,v)](b_k)=x_{k+v+1},
\end{gather*}
where $v\in\N$, and $\bot_D=\top_D=\{x_0\}$.
This is similar to the translation of a $1$-VASS 
\[
(\{x\},\{(x,1,x),(x,-1,x)\})
=
\vcenter{\hbox{\begin{tikzpicture}
\node[state, minimum size=10pt] (x) at (0,0) {$x$};
\path (x) edge[loop left, looseness=30] node[left] {$(1)$} node[above right=3pt] {$a$} (x);
\path (x) edge[loop right, looseness=30] node[right] {$(-1)$} node[above left=3pt] {$b$}  (x);
\end{tikzpicture}}}
\]
in Example \ref{x:VASStoCounter}; the difference being that the transitions are labelled by $a$ and $b$, respectively. 

Proposition \ref{p:LanguagesOfCounters} shows that the categories of $(\cG\times\N)$-words and $\cG$-words are isomorphic. Thus, $\Words_{(\vo,\star)}(\cG\times \N)\cong \Words_{\vo}(\cG)\cong \{a,b\}^*$.
Every accepting track in $D$ corresponds to an accepting run of this VASS, as in Example \ref{x:VASStoCounter}. So $\Lang(D)\subseteq \{a,b\}^*$ is the Dyck language: it consists of all words with an equal number of letters $a$ and $b$, and such that the number of $a$'s in every prefix is no less than the number of $b$'s.

The automaton $D$ has finite type, as it is isomorphic to the colimit of the following diagram of representable presheaves:
\[
\begin{tikzcd}[row sep=tiny,column sep=large]
    & \widehat{(\vo,\star)}
        \ar[dd,"{(\id_\vo,1)}"]
        \ar[dl,"{(\tau_a,0)}",swap]
        \ar[dr,"{(\sigma_b,0)}"]
    & \\
    \widehat{(a,\star)} & & \widehat{(b,\star)}\\
    & \widehat{(\vo,\star)}
        \ar[ul,"{(\sigma_a,0)}"]
        \ar[ur,"{(\tau_b,0)}",swap]
& 
\end{tikzcd}
\]
An isomorphism of the colimit of this diagram with $D$ is obtained by maps given by the elements $a_0$, $b_0$, $x_0$ and $x_1$ for the the left, right, lower and upper object, respectively.
A generalisation of this construction is presented in the next section.
\end{exa}

\begin{prp}
	Kleene's theorem fails for $(\cG(\{a,b\})\times \N)$-automata.
\end{prp}
\begin{proof}
The $(\cG\times\N)$-automaton $D$ from Example~\ref{x:Dyck} shows that the Dyck language is $(\cG\times \N)$-regular: $\Lang(D)\in \Reg(\cG\times\N)$. But it is not $\cG$-regular: $\Lang(D)\not\in I_*(\Reg(\cG))$. Thus, by Proposition \ref{p:LanguagesOfCounters} and 
    Kleene's  theorem for standard automata,
$
I_*(\Reg(\cG))=I_*(\Rat(\cG))=\Rat(\cG\times\N)\subsetneq \Reg(\cG\times \N).
$    
\end{proof}

\section{Higher-dimensional automata with counters}
\label{s:HDAC}

We conclude with introducing higher-dimensional automata with
counters, which generalise both higher-dimensional automata and
VASSes.

Fix an alphabet $\Sigma$.  Let
$\square(\Sigma)\subseteq \CList(\Sigma)$ be the skeleton of the
d-category from Section \ref{s:HDA} with
\begin{itemize}
\item
objects 
$[n,\mu]=([n],\mu:[n]\to\Sigma)$,
where $[n]=\{1\intord\dotsm\intord n\}$, and
\item
	generating morphisms $d^\alpha_i:[n-1,\mu\circ \delta_i]\to[n,\mu]$
as in \eqref{e:DeltaEpsilon}.
\end{itemize}
Recall that $\square(\Sigma)$-presheaves and
$\square(\Sigma)$-automata are precubical sets and higher-dimensional
automata of \cite{FJSZ_HDALang,FahrenbergJSZ24}, respectively.

\begin{df}
  A \emph{higher dimensional automaton with $r$ counters}
  (\emph{HDAC}) is a presheaf automaton over $\sq(\Sigma)\times \N^r$.
\end{df}

The languages of HDAC and HDA are equivalent owing to Proposition~\ref{p:LanguagesOfCounters}.  Languages of HDA are studied in~\cite{FJSZ_HDALang,FahrenbergJSZ24}.

Van Glabbeek has shown that (infinite) HDA are more expressive than
Petri nets \cite{vanGlabbeek_Expressiveness}.  His construction
produces an HDA whose accessible part is finite only if the Petri net
is safe.  Here we show that  HDACs of finite type subsume all Petri nets
-- with the safeness assumption lifted.

Let $\cP=(P,T,F)$ be a labelled Petri net with places $P$, transitions
$T$, flows $F\subseteq (P\times T)\sqcup(T\times P)$ and a labeling
function $\lambda:T\to\Sigma$.  Consider the functions
$\sfa_0,\sfa_1:T\to\{0,1\}^P$ given by
\[
  \sfa_0(t)_p=
  \begin{cases}
    1 & \text{if } (p,t) \in F,\\
    0 & \text{if } (p,t)\not\in F,
  \end{cases}
  \qquad
  \sfa_1(t)_p=
  \begin{cases}
    1 & \text{if } (t,p)\in F,\\
    0 & \text{if } (t,p)\not\in F.
  \end{cases}
\]
For $\bt=(t_1,\dotsc,t_n)\in T^n$
we write
\[
  \lambda\bt=[n, i \mapsto \lambda(t_i)]\in\square(\Sigma),
  \qquad
  \partial_i\bt =(t_1,\dotsc,t_{i-1},t_{i+1},\dotsc,t_n)\in T^{n-1}.
\]
The HDAC $X=X_\cP\in\aut{\sq(\Sigma)\times \N^P}$ associated with the
Petri net $\cP$ is defined by
\begin{align*}
  X[([n,\mu],\star)]&= \{(\bt,\sfv)\in T^n\times \N^P\mid
                      \lambda\bt=[n,\mu]\},\\
  X[(d^\varepsilon_i,\sfw)](\bt;\sfv)&=(\partial_i\bt;\sfv+\sfa_\varepsilon(t_i)+\sfw).
\end{align*}

This is essentially the construction in \cite{vanGlabbeek_Expressiveness}, the only difference being the enrichment of the underlying category with labels.  In general, $X$ is not of finite type, but a slight modification of $X$ has this property.

Let $X_{nac}$ and $X_{{\leq}d}$ for $d\leq 1$ be the sub-automata of
$X$ given by
\begin{align*}
  X_{nac}[([n,\mu],\star)]&=\{(t_1,\dotsc,t_n;\sfv) \in X[([n,\mu],\star)] \mid \forall i\neq j.\; t_i\neq t_j \},\\
  X_{{\leq}d}[([n,\mu],\star)]&=\{(t_1,\dotsc,t_n;\sfv) \in X[([n,\mu],\star)] \mid n\leq d \}.
\end{align*}
In $X_{nac}$ we restrict to collections of different transitions (that
is, there is no autoconcurrency), and in $X_{{\leq}d}$ to collections
of at most $d$ parallel transitions.

\def\sh{\mathsf{sh}} We start with a presentation of $X$ which is not
finite, but specialises to finite presentations of
 $X_{nac}$ and $X_{\leq d}$. It generalises the presentation in Example \ref{x:Dyck}.
 
 Let $\cE$ be the category with objects
\[
	\Ob(\cE)=\left(\coprod_{n\geq 0} T^n\right) \cup \left( \coprod_{n\geq 0} T^n\times \{1,\dotsc,n\}\times\{0,1\}\right)
\]
and non-identity morphisms 
\[
	\varphi_{(\bt;i,\varepsilon)}: (\bt;i,\varepsilon)\to\partial_i\bt ,
	\qquad
	\psi_{(\bt;i,\varepsilon)}: (\bt;i,\varepsilon)\to \bt,
\]
for $n\geq 0$, $\bt\in T^n$, $1\leq i\leq n$ and $\varepsilon\in\{0,1\}$.
Composition is trivial.  Let $F:\cE\to\Cell X$ be the functor defined by
\begin{equation}
\label{e:XPPresMap}
F(\bt)=((\lambda\bt,\star), (\bt,\sfO))
,\qquad
F(\bt;i,\varepsilon)=((\lambda\partial_i\bt,\star),(\partial_i\bt,\sfa_\varepsilon(t_i))
\end{equation}
on objects and  by           
\begin{equation*}
  F(\varphi_{(\bt;i,\varepsilon)})
  =
  ((\id_{\lambda\partial_i\bt},\sfa_\varepsilon(t_i)),(\partial_i\bt,\sfO)),
    \qquad
  F(\psi_{(\bt;i,\varepsilon)})
  =
  ((d^\varepsilon_i,\sfO),(\bt,\sfO))
\end{equation*}
on morphisms. Let $G=\pi_X\circ F:\cE\to\square(\Sigma)\times\N^P$, so that values
of $G$ are the first coordinates of the values of $F$ in the formulas
above. The functor $F$ determines the presheaf map \begin{equation*}
    f:\widehat{G}\xrightarrow{F_*} \widehat{\Cell X} \cong X
\end{equation*}
induced by the cocone $(\widehat{G(e)}\xrightarrow{F(e)} X)_{e\in\cE}$, via the identification \eqref{e:Yoneda}.

\begin{lem}
\label{l:XPPres}
	The presheaf map $f$ is an isomorphism.
\end{lem}
\begin{proof}
  For $e\in\cE$, let $I_{e}:\widehat{G(e)}\to\widehat{G}$ denote the canonical injection. Since $(I_e)_{e\in\cE}$ is a cocone, $I_e\circ (\widehat{G(e')}\xrightarrow{G(\alpha)}\widehat{G(e)})=I_{e'}$ for any morphism $\alpha\in\cE(e',e)$, that is,
  \[
    I_{e'}(x)=I_e(G(\alpha)\circ x)
  \]
  for $U\in\square(\Sigma)$ and $x\in \widehat{G(e)}[(U,\star)]=(\square(\Sigma)\times \N^r)((U,\star),G(e))$.
  In particular,
  \[
    I_{(\bt;i,\varepsilon)}(\omega, \sfv)=
    I_{\partial_i\bt}(\omega,\sfv+\sfa_\varepsilon(t_i))
    \qquad\text{and}\qquad
    I_{(\bt;i,\varepsilon)}(\omega, \sfv)=
    I_{\bt}(d^\varepsilon_i\circ\omega,\sfv)
  \]
  for $\alpha=\varphi_{(t;i,\varepsilon)}$ and $\alpha=\psi_{(\bt;i,\varepsilon)}$, respectively.
  For $\bt\in T^n$ and $\sfv\in\N^P$, we define
  \[
    g(\bt,\sfv)=I_{\bt}(\id_{\lambda\bt},\sfv)\in\widehat{G}[(\lambda\bt,\star)].
  \]
  Note that ($\id_{\lambda\bt},\sfv)$ is an endomorphism of   $(\lambda\bt,\star)$ and therefore an element of   $\widehat{(\lambda\bt,\star)}[(\lambda\bt,\star)]=\widehat{G(\bt)}[(\lambda\bt,\star)]$.
  For every $(\bt,\sfv)\in X[(\lambda\bt,\star)]$ and morphism
  $(d^0_i,\sfw): (\lambda\partial_i\bt,\star)\to (\lambda\bt,\star)$
  we thus have
  \begin{align*}
    g(X[(d^0_i,\sfw)](\bt,\sfv))
    &= g(\partial_i\bt,\sfv+\sfa(t_i)+\sfw)\\
    &=I_{\partial_i\bt}(\id_{\lambda\partial_i\bt},\sfv+\sfa(t_i)+\sfw)\\
    &=   I_{(\bt;i,0)}(\id_{\lambda\partial_i\bt},\sfv+\sfw)\\
&=I_{\bt}(d^0_i:\partial_i\lambda\bt\to\lambda\bt,\sfv+\sfw)\\
  &=\widehat{G}[(d^0_i,\sfw)](I_{\bt}(\id_{\lambda\bt},\sfv))\\
&=    \widehat{G}[(d^0_i,\sfw)](g(\bt,\sfv)).
  \end{align*}
  This show that $g$ is a presheaf map $X\to\widehat{G}$.
  Furthermore, $f\circ g=\id_{X}$ since
  \[
    f(g(\bt,\sfv))
    =
    f(I_\bt(\id_{\lambda\bt},\sfv))
    =
    X[(\id,\sfv)]f(I_\bt(\id_{\lambda\bt},\sfO))
    =
    X[(\id_{\lambda\bt},\sfv)](\bt,\sfO)
    =
    (\bt,\sfv).
  \]
  Finally, $g$ is surjective because $\widehat{G}$ is spanned by the elements
  $I_\bt(\id_{\lambda\bt},\sfO)=g(\bt,\sfO)$.
  This shows that $g$ is an inverse of $f$; so they are isomorphisms.
\end{proof}

\begin{prp}
	The HDACs $X_{nac}$ and $X_{{\leq}r}$, for $r\geq 1$, have finite type.
\end{prp}
\begin{proof}
  Let $\cE_{nac}$ and $\cE_{\leq r}$ be the full subcategories of
  $\cE$ with objects $\bt=(t_1,\dotsc,t_n)$ and
  $(\bt;i,\varepsilon)$, such that either the
  $\lambda(t_i)$'s are pairwise different (for $X_{nac}$) or $n\geq r$
  (for $X_{{\leq}r}$).
Since the set $T$ of transitions is finite, $\cE_{\leq r}$ is a finite category and $\cE_{nac}\subseteq\cE_{\leq r}$ whenever $|T|\leq r$.  
   The restriction of $F$ in \eqref{e:XPPresMap} to
  $\cE_{nac}$ or $\cE_{\leq r}$ gives finite presentations of
  $X_{nac}$ and $X_{{\leq}r}$, respectively, by the same argument as in the proof of Lemma \ref{l:XPPres}.
\end{proof}

\section{Conclusion}

We have introduced presheaf automata as a generalisation of different
variants of higher-dimensional automata, introducing in particular
d-categories as a generalisation of the index categories that generate
them. We have shown that standard automata can be modelled as presheaf
automata as well, and likewise, as a more substantial contribution, Petri nets
without a safety assumption. We have also outlined the foundations of
a language theory for presheaf automata, including definitions of
regular and rational languages and a notion of open map for presheaf
automata, which forms the basis for notions of bisimulations. These
concepts may lead to analoga of classical theorems for automata, such
as Kleene's theorem, the Myhill-Nerode theorem or the
B\"uchi-Elgot-Trakhtenbrot theorem for some instances of presheaf
automata, extending recent work on higher-dimensional
automata~\cite{ABFM_Logic,FahrenbergJSZ24,FZ_MyhillNerode-Conf}. However, our results rule out a generic Kleene theorem for all presheaf automata.

There is a long history of categorical approaches to standard automata
and a vast literature on the topic, starting, to our knowledge, with
work by Eilenberg and Wright
\cite{Eilenberg_Wright-AutomtaInGeneralAlgebras}, and both from an
algebraic and a coalgebraic point of view, as the following
representative texts show~\cite{Adamek-Trnkova_Automata,Rutten00}. A
recent functorial approach to standard automata has been described by
Colcombet and Petrisan~\cite{ColcombetP19}, who
cite~\cite{Bainbridge74} as an early predecessor. A presheaf approach
to standard automata has been proposed by
Rosenthal~\cite{Rosenthal95}, but using relational presheaves, which
are lax functors $\cC^{\textup{op}}\to \mathbf{Rel}$ into the category
of sets and binary relations. Both approaches have little relationship
to ours. Katis et al.~\cite{KatisSW97} model transition systems as
spans in the category of graphs, which is reminiscent of the
track objects introduced in this article, yet the construction of a d-category for their setting does not seem obvious and is left for future work. Melliès and Zeilberger~\cite{MelliesZ25} model automata over categories and operators using a fibrational approach; their objective to model not only standard automata seem similar to ours, but instead of using presheaves, their automata are essentially functors into a category.

There is also a long line of work on transition systems and Petri nets
as presheaves, starting
with~\cite{Joyal-Winskel-Nielsen_Bisimulation}. The seminal article of
Meseguer and Monanari \cite{Meseguer-Montanari_PetriNetsAreMonoids} is
similar in spirit to our examples towards the end of this article, but it emphasises
symmetric monoidal categories, which are not explicit in our work,
while presheaves are missing. Formalising cubical sets as presheaves,
as well as their simplicial or globular counterparts, is standard in
mathematics. Van Glabbeek \cite{vanGlabbeek_Expressiveness} and
subsequently Goubault and Mimram \cite{Goubault-Mimram_Relationship}
have previously related higher-dimensional automata with other models
of concurrency, including transition systems, event structures or
Petri nets. By contrast, all comparisons made in previous sections
are original, to our knowledge. 

Beyond the language-theoretic perspective mentioned above, we envisage
two main avenues for future work. Obviously, d-functors induce natural
transformations between presheaves on the source and target
d-categories. This provides tools for translating between different
instances of presheaf automata. A natural question asks for conditions
on d-functors that preserve various properties of presheaf automata.
Further, the relation between presheaf automata and monoidal
structures in concurrency, as for instance in
\cite{Meseguer-Montanari_PetriNetsAreMonoids}, remains to be explored,
see also the work of Grandis and Mauri for a monoidal approach to
cubical sets~\cite{Grandis-Mauri_Cubical}, the work of Sassone and
Soboci{\'n}ski and subsequently Baez and Master on open Petri nets in
the setting of symmetric monoidal bicategories and double
categories~\cite{SassoneS05,BaezM20}, as well as more recent work on
Petri nets by Baez et
al~\cite{BaezGMS21}. Soboci{\'n}ski uses
  relational presheaves for modelling structured labels in transition
  systems~\cite{Sobocinski15}. It might thus be interesting to consider our
  approach in this context. An alternative interesting
  approach to structured labels in Petri nets via groupoids has been
  introduced by Kock~\cite{Kock23}. Finally, see \cite{FJSZ_HDALang}
for a definition of parallel composition of higher-dimensional
automata as a tensor.

\paragraph{Acknowledgement}
The authors are grateful to Uli Fahrenberg and Christian Johansen for fruitful discussions and helpful feedback on this paper.

\bibliographystyle{plain}
\bibliography{kzbib}

\begin{thebibliography}{10}

\bibitem{Adamek-Trnkova_Automata}
Jiří Adámek and Věra Trnková.
\newblock {\em Automata and algebras in categories}.
\newblock Mathematics and its applications. Kluwer Acad. Publ., 1990.

\bibitem{ABFM_Logic}
Amazigh Amrane, Hugo Bazille, Uli Fahrenberg, and Marie Fortin.
\newblock {L}ogic and {L}anguages of {H}igher-{D}imensional {A}utomata.
\newblock In Joel~D. Day and Florin Manea, editors, {\em {DLT} 2024}, volume
  14791 of {\em Lect. Notes Comput. Sci.}, pages 51--67. Springer, 2024.

\bibitem{Antolini00}
Rosa Antolini.
\newblock Cubical structures, homotopy theory.
\newblock {\em Ann. Mat. Pura Appl.}, 178(4):317--324, 2000.

\bibitem{BaezGMS21}
John~C. Baez, Fabrizio Genovese, Jade Master, and Michael Shulman.
\newblock Categories of nets.
\newblock In {\em {LICS} 2021}, pages 1--13. {IEEE}, 2021.

\bibitem{BaezM20}
John~C. Baez and J.~Master.
\newblock Open {P}etri nets.
\newblock {\em Math. Struct. Comput. Sci.}, 30(3):314--341, 2020.

\bibitem{Bainbridge74}
E.~S. Bainbridge.
\newblock Adressed machines and duality.
\newblock In Ernest~G. Manes, editor, {\em Category Theory Applied to
  Computation and Control 1974}, volume~25 of {\em Lect. Notes Comput. Sci.},
  pages 93--98. Springer, 1974.

\bibitem{ChomskyS63}
N.~Chomsky and M.~P. Sch{\"u}tzenberger.
\newblock The algebraic theory of context-free languages.
\newblock In {\em Computer programming and formal systems}, pages 118--161.
  North-Holland, 1963.

\bibitem{ColcombetP19}
Thomas Colcombet and Daniela Petrişan.
\newblock Automata minimization: a functorial approach.
\newblock {\em Log. Methods Comput. Sci.}, 16(1), 2020.

\bibitem{Crans95}
Sjoerd~Erik Crans.
\newblock {\em On Combinatorial Models for Higher Dimensional Homotopies}.
\newblock PhD thesis, Universiteit Utrecht, 1995.

\bibitem{Eilenberg_Wright-AutomtaInGeneralAlgebras}
Samuel Eilenberg and Jesse~B. Wright.
\newblock Automata in general algebras.
\newblock {\em Inform. Control}, 11(4):452--470, 1967.

\bibitem{FJSZ_HDALang}
Uli Fahrenberg, Christian Johansen, Georg Struth, and Krzysztof Ziemia{\'n}ski.
\newblock Languages of higher-dimensional automata.
\newblock {\em Math. Struct. Comput. Sci.}, 31(5):575--613, 2021.

\bibitem{FahrenbergJSZ24}
Uli Fahrenberg, Christian Johansen, Georg Struth, and Krzysztof Ziemia{\'n}ski.
\newblock Kleene theorem for higher-dimensional automata.
\newblock {\em Log. Methods Comput. Sci.}, 20(4), 2024.

\bibitem{FZ_MyhillNerode-Conf}
Uli Fahrenberg and Krzysztof Ziemiański.
\newblock A {M}yhill-{N}erode {T}heorem for {H}igher-{D}imensional {A}utomata.
\newblock In Lu{\'{\i}}s Gomes and Robert Lorenz, editors, {\em {PETRI} {NETS}
  2023}, volume 13929 of {\em Lect. Notes Comput. Sci.}, pages 167--188.
  Springer, 2023.

\bibitem{FGHMR_DAT}
Lisbeth Fajstrup, Eric Goubault, Emmanuel Haucourt, Samuel Mimram, and Martin
  Raussen.
\newblock {\em Directed {A}lgebraic {T}opology and {C}oncurrency}.
\newblock Springer, 2016.

\bibitem{Goubault-Mimram_Relationship}
Éric Goubault and Samuel Mimram.
\newblock Formal relationships between geometrical and classical models for
  concurrency.
\newblock {\em Electron. Notes Theor. Comput. Sci.}, 283:77--109, 2012.

\bibitem{Grandis-Mauri_Cubical}
Marco Grandis and Luca Mauri.
\newblock Cubical sets and their site.
\newblock {\em Theor. Appl. Categ.}, 11:185--211, 2003.

\bibitem{Hopcroft-Pansion_Reachability}
John Hopcroft and Jean-Jacques Pansiot.
\newblock On the reachability problem for 5-dimensional vector addition
  systems.
\newblock {\em Theor. Comput. Sci.}, 8(2):135--159, 1979.

\bibitem{Jardine02}
J.~F. Jardine.
\newblock Cubical homotopy theory: a beginning.
\newblock Technical Report NI02030-NST, Newton Institute, 2002.

\bibitem{Johnson-Yau_2Categories}
Niles Johnson and Donald Yau.
\newblock {\em 2-Dimensional Categories}.
\newblock Oxford University Press, 2021.

\bibitem{Joyal-Winskel-Nielsen_Bisimulation}
André Joyal, Mogens Nielsen, and Glynn Winskel.
\newblock Bisimulation from open maps.
\newblock {\em Inform. Comput.}, 127(2):164--185, 1996.

\bibitem{Kan55}
Daniel~M. Kan.
\newblock Abstract homotopy. {I}.
\newblock {\em Proc. Natl. Acad. Sci. USA}, 41:1092--1096, 1955.

\bibitem{Kashiwara-Schapira_Categories}
Masaki Kashiwara and Pierre Schapira.
\newblock {\em Categories and Sheaves}.
\newblock Springer, 2006.

\bibitem{KatisSW97}
Piergiulio Katis, Nicoletta Sabadini, and Robert F.~C. Walters.
\newblock {S}pan({G}raph): {A} {C}ategorial {A}lgebra of {T}ransition
  {S}ystems.
\newblock In Michael Johnson, editor, {\em {AMAST} '97}, volume 1349 of {\em
  Lect. Notes in Comput. Sci.}, pages 307--321. Springer, 1997.

\bibitem{Kock23}
Joachim Kock.
\newblock Whole-grain {P}etri {N}ets and {P}rocesses.
\newblock {\em J. {ACM}}, 70(1):1:1--1:58, 2023.

\bibitem{Loregian-Riehl_Categorical}
Fosco Loregian and Emily Riehl.
\newblock Categorical notions of fibration.
\newblock {\em Expo. Math.}, 38(4):496--514, 2020.

\bibitem{MacLane_Categories}
Saunders Mac~Lane.
\newblock {\em Categories for the Working Mathematician}.
\newblock Springer, 1978.

\bibitem{MacLane-Moerdijk_Sheaves}
Saunders Mac~Lane and Ieke Moerdijk.
\newblock {\em Sheaves in Geometry and Logic: A First Introduction to Topos
  Theory}.
\newblock Springer, 1994.

\bibitem{MelliesZ25}
Paul-Andr{\'e} Melli{\`e}s and Noam Zeilberger.
\newblock The {C}ategorical {C}ontours of the {C}homsky-{S}ch{\"u}tzenberger
  {R}epresentation {T}heorem.
\newblock {\em Log. Methods Comput. Sci.}, 21(2):12:1--12:49, 2025.

\bibitem{Meseguer-Montanari_PetriNetsAreMonoids}
José Meseguer and Ugo Montanari.
\newblock Petri nets are monoids.
\newblock {\em Inform. Comput.}, 88(2):105--155, 1990.

\bibitem{Pratt_ModelingConcurrency}
Vaughn Pratt.
\newblock Modeling concurrency with geometry.
\newblock In {\em POPL ’91}, POPL ’91. ACM Press, 1991.

\bibitem{Rosenthal95}
Kimmo~I. Rosenthal.
\newblock Quantaloids, enriched categories and automata theory.
\newblock {\em Appl. Categorical Struct.}, 3(3):279--301, 1995.

\bibitem{Rutten00}
Jan J. M.~M. Rutten.
\newblock Universal coalgebra: a theory of systems.
\newblock {\em Theor. Comput. Sci.}, 249(1):3--80, 2000.

\bibitem{SassoneS05}
Vladimiro Sassone and Pawe{\l} Soboci{\'n}ski.
\newblock A {C}ongruence for {P}etri nets.
\newblock In Hartmut Ehrig, Julia Padberg, and Grzegorz Rozenberg, editors,
  {\em PNGT@ICGT 2004}, volume 127 of {\em Electron. Notes Theor. Comput.
  Sci.}, pages 107--120. Elsevier, 2004.

\bibitem{Serre51}
Jean-Pierre Serre.
\newblock Homologie singuli{\`e}re des espaces fibr{\'e}s. {A}pplications.
\newblock {\em Ann. of Math. (2)}, 54:425--505, 1951.

\bibitem{Sobocinski15}
Pawe{\l} Soboci{\'n}ski.
\newblock Relational presheaves, change of base and weak simulation.
\newblock {\em J. Comput. Syst. Sci.}, 81(5):901--910, 2015.

\bibitem{vanGlabbeek_Expressiveness}
R.~J. van Glabbeek.
\newblock On the expressiveness of higher dimensional automata.
\newblock {\em Theor. Comput. Sci.}, 356(3):265--290, 2006.

\end{thebibliography}

\end{document}